\documentclass[aps,pra,superscriptaddress,twocolumn,showpacs]{revtex4-1}
\usepackage{bm} \usepackage{graphicx} \usepackage{amsmath}
\usepackage{amsthm,stmaryrd}
\usepackage{amssymb} 
\usepackage{txfonts}
\usepackage{color}
\usepackage{xcolor}
\usepackage{subfigure}

\newcommand{\ket}[1]{|{#1}\rangle}
\newcommand{\bra}[1]{\langle{#1}|}

\newcommand{\elmat}[3]{\langle #1\vert #2 \vert #3 \rangle}
\newcommand{\pr}{\mathbb{P}}
\newcommand{\pq}{\mathbb{Q}}
\newcommand{\proj}[2]{\vert #1\rangle \langle #2 \vert}

\DeclareRobustCommand\openzero{\leavevmode\hbox{0\kern-.55em0}}
\mathchardef\minus="002D

\begin{document}
\title{NOON States via Quantum Walk of Bound Particles}

\author{Enrico Compagno}
\affiliation{Department of Physics and Astronomy, University College London, Gower Street, WC1E 6BT London, United Kingdom}

\author{Leonardo Banchi}
\affiliation{Department of Physics and Astronomy, University College London, Gower Street, WC1E 6BT London, United Kingdom}

\author{Christian Gross}
\affiliation{Max-Planck-Institut f\"{u}r Quantenoptik, 85748 Garching, Germany}

\author{Sougato Bose}
\affiliation{Department of Physics and Astronomy, University College London, Gower Street, WC1E 6BT London, United Kingdom}

\date{\today}
\begin{abstract}
Tight-binding lattice models allow the creation of bound composite objects
which, in the strong-interacting regime,  are protected against dissociation. We
show that a local impurity in the lattice potential can generate a coherent
split of an incoming bound particle wave-packet which consequently produces a
NOON state between the endpoints. This is non trivial because when finite
lattices are involved, edge-localization effects make their use for
non-classical state generation and information transfer challenging. We derive
an effective model to describe the propagation of bound particles 
in a Bose-Hubbard chain. We introduce local impurities in the lattice potential
to inhibit localization effects and to split the propagating bound particle,
thus enabling the generation of distant NOON states. We analyze how minimal engineering transfer schemes improve the transfer fidelity and we quantify the robustness to typical decoherence effects in optical lattice  implementations.  Our scheme potentially has an impact on quantum-enhanced atomic interferometry in a lattice.
\end{abstract}

\maketitle

\section{Introduction}
The unprecedented ability to control and observe multi-particle states in 
optical lattice systems with single-site resolution 
\cite{weitenberg_single-spin_2011,preiss_strongly_2015,fukuhara_microscopic_2013,fukuhara_quantum_2013,murmann_two_2015,preiss_quantum_2015,robens_quantum_2015,celi_synthetic_2014,mancini_observation_2015,chiara_detection_2011,islam_measuring_2015,zupancic_ultra-precise_2016,ott_single_2016,hoffmann_holographic_2016}
make possible the investigation of new quantum interference effects. 
Indeed, the dynamics of quantum interacting systems display many interesting
features that go beyond the regime traditionally studied in linear optics. 
From the fundamental perspective it is then important to understand how to
exploit the natural interactions to ``engineer'' the  many-particle dynamics 
in a lattice for creating non-classical states, such as multi-particle NOON
states. Compared to classical setups, and also to other schemes for atom interferometry \cite{rocco_fluorescence_2014,kovachy_quantum_2015}, the advantage of this approach is that
non-classical states 
(e.g. NOON and dual Fock states \cite{lee_quantum_2002})
enhance the estimation precision of the phase difference between the output arms of an interferometer \cite{dowling_quantum_2008,demkowicz-dobrzanski_chapter_2015,gross_nonlinear_2010,gross_spin_2012}, making them highly attractive for technological applications. Super-resolution for NOON states with $N=2,3$ has been recently shown experimentally for microscopy purposes \cite{israel_supersensitive_2014}.
However, the generation of non-classical states with high-fidelity 
is still a hard task. For instance, 
in existing photonic realizations, NOON states with $N=5$ have been
demonstrated, but with a limited $42\%$ fringe visibility
\cite{hofmann_high-photon-number_2007,afek_high-noon_2010,israel_experimental_2012}.
Moreover, with those schemes, there is a theoretical upper 
threshold for the state
preparation fidelity of $94.3\%$ \cite{hofmann_high-photon-number_2007}.
It is therefore important to develop alternative schemes for high-fidelity 
NOON states generation.

To sense spatial inhomogeneities and to probe external fields localized over few sites, it is convenient for the components of the NOON state to be spatially well separated. In this context a quantum walk of interacting atoms in an optical lattice might be very useful as we shall explore. For a lattice setup this type of scheme is important as it enables one to avoid the necessity of measurement based schemes \cite{kok_creation_2002,mullin_creation_2010} (which are still 
challenging in current optical lattice experiments with few particles), 
time-dependent external potentials \cite{stiebler_spatial_2011},
engineered bath based schemes \cite{kordas_dissipation-induced_2012} or ring lattices
\cite{leung_dynamical_2012,dellanna_entanglement_2013}. From the theoretical point of view, optical lattice systems in a low filling limit are modeled by the 
Bose-Hubbard Hamiltonian, which contains a hopping term between neighboring
sites and an onsite interaction. 
If more than one particles are initially located in the same site and the onsite interaction is sufficiently
strong, this model favors the creation of bound states, which are stable against dissociation \cite{winkler_repulsively_2006,petrosyan_quantum_2007,folling_direct_2007,valiente_two-particle_2008,valiente_three-body_2010,preiss_strongly_2015}. A natural question is then whether bound
states have some advantages for non-classical state production tasks. The key
point here is that a bound state behaves like an effective single particle for strong enough interaction and together with a balanced beam splitter it can be used to produce a high fidelity NOON state.
When distant sites are involved, a balanced beam splitter transformation can be obtained via the particle dynamics by introducing suitable impurities in the lattice potential. These
impurities generate a coherent splitting of the wave-packet of propagating
particles which enables high-efficiency effective linear optical operations
between remote sites of finite lattices
\cite{compagno_toolbox_2015,banchi_perfect_2015}. Even without the onsite
interaction, peculiar quantum interference effects enable the production of 
non-classical states, namely two particle NOON states, via the celebrated 
Hong-Ou-Mandel effect \cite{compagno_toolbox_2015,banchi_perfect_2015}.  
Being a linear-optical effect, 
the efficiency of this protocol is maximized when the atom-atom interaction is kept in the weak coupling regime.

On the other hand it is intriguing to investigate the strongly interacting
regime, namely whether the role of the inter-atomic interaction can be exploited
as a resource to generate a wider class of non-classical states, for instance
high NOON states ($N>2$) between distant sites. 
\begin{figure}[t]
\centering
\includegraphics[trim={0cm 0cm 0cm 1cm},clip=True,width=1\columnwidth]{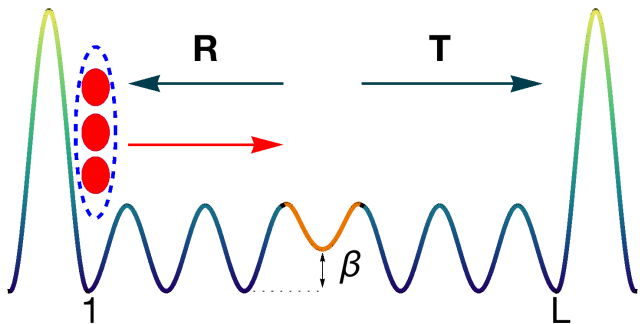}
\caption{Scheme of the model: when a bound state is initially in the site $1$
    of a finite lattice, a suitably introduced local impurity (orange) in the
    chemical potential $\mu_{j}=-\beta \delta_{j,L/2+1}$ triggers a wave-packet
splitting in a reflected $R$ and a transmitted $T$ component. If appropriately
tuned it generates a NOON state between the endpoints (site $1$ and $L$) at the
transfer time. The two green peaks are local fields which realize a finite
lattice model. The red arrow represents the natural direction of propagation of
the bound particle once lowered the lattice depth at time $t=0$. We also add
local fields in the first and last sites of the chain to inhibit the 
edge-localization effect and then to delocalize the bound particle from the edges.}
\label{fig:BPUScheme}
\end{figure}
However, as far as distant sites are concerned, 
the realizability of this scheme is hindered by the possibility to engender and
control the tunneling dynamics of a bound state initially located in one edge of
the lattice. The main obstacle is the presence of edge-locked states which
inhibit the hopping dynamics
\cite{pinto_edge-localized_2009,haque_self-similar_2010}. Edge-locking indeed
creates an effective energy barrier between edge and bulk sites, that suppresses
the bound state propagation along the lattice.

In other words, it is still an open problem how to tune the lattice potential to
realize transformations between far sites when strongly interacting
particles are involved. 
Recently, in the case of fermions, a long-range state transfer protocol for a
two particle bound state has been studied in a one- and two-dimensional
lattices, using AC fields. In this scheme the state transfer takes
place only between edge states, while bulk sites remain empty during the
dynamics of the system \cite{bello_long-range_2016,bello_sublattice_2016}. 
On the other hand, for bosonic particles, the under-barrier tunneling of a 
dimer has been
analyzed in \cite{kolovsky_energetically_2012,maksimov_escape_2014}. 
Nonetheless, the problem of how to transfer bound states with high fidelity over
arbitrarily long distances in engineered finite-size chains has not been completely
addressed yet, in particular when more than two particle bound states 
are involved in the dynamics.

In this paper we analyze the bound particle dynamics in a finite lattice by mapping
the Bose-Hubbard Hamiltonian into an effective single particle chain via a
strong-coupling expansion in the onsite interaction term. This mapping is 
realized with a novel application of the effective theory developed in 
\cite{jia_integrated_2015}.
We study the conditions to prevent the dissociation of the bound particle during
the dynamics and we show that when these conditions are satisfied the bound
particle evolution is perfectly described by our effective model.  We
find the connection between the effective hopping rates, which are interaction
dependent, and the physical parameters of the of the Bose-Hubbard model. We then
show how to design these parameters such that the effective evolution produces a
splitting transformation suitable for creating NOON states between distant
sites. Applications in quantum enhanced metrology are thus discussed.  

The first step towards the realization of our protocol is to develop a method to
delocalize a bound state wave-packet from the edges of a finite chain. In the
spin chain case the edge-locking effect for spin blocks is bypassed using $\pi$
pulses  to flip the leftmost spin and then enable the wave-packet delocalization
\cite{haque_self-similar_2010,alba_bethe_2013}. On the other hand, 
in Bose-Hubbard model an operative method to unlock bound particle states 
has not been proposed yet.
Here we show that the edge locking effect can be eliminated by introducing
static impurities in the chemical potential localized at the endpoints. These
impurities, which can be generated using external local fields, compensates 
the energy gap between edge and bulk sites and enable the dynamics. 
After ``unlocking'' the dynamics, 
we study the state transfer efficiency for a uniform chain and the robustness
from typical environmental effects, specifically decoherence effects due to
spontaneous emission in an optical lattice setup. Moreover we show how a minimal
engineering of the hopping rates can enhance the transfer efficiency, 
specifically tuning the first and the last tunneling couplings of the chain.
We show then how to add an extra impurity in the middle of the chain to generate a  NOON state between the edges of the lattice. 
We derive analytical expressions for the optimal parameters to generate $N=2$
and $N=3$ NOON states,
and we show how our approach can be straightforwardly generalized for producing
larger ``cat'' states. 
Finally, we show how to experimentally detect the generated NOON state 
using the technology available nowadays. 

\section{Main idea}
We consider a one-dimensional chain of length $L$, described by a Bose-Hubbard
model with site dependent parameters according to the following Hamiltonian
\cite{lewenstein_ultracold_2012}: 
\begin{equation}
H=-\sum\limits_{j=1}^{L-1}\frac{J_j}{2}\left[a_j a_{j+1}^\dag + H.c.
\right]+\sum\limits_{j=1}^{L} \frac{U_j}{2}
n_j\left(n_j+1\right)-\sum\limits_{j=1}^{L}\mu_j n_j~.
\label{eq:BoseHubbard}
\end{equation}
Here $a_j (a_j^\dag)$ are the boson annihilation (creation) operators,
$n_j=a_j^\dag a_j$ the number operator and $J_j$, $U_j$ and $\mu_j$ are
respectively the hopping rate, the onsite interaction and the chemical
potential. Because the Hamiltonian, Eq. \eqref{eq:BoseHubbard}, preserves the total
number of excitations of the system, the dynamics can be evaluated in a Hilbert
subspace with a fixed number of particles. 

One characteristic feature of the Bose-Hubbard model is that the onsite
interaction enables the creation of ``bound'' states when several particles
are in the same site
\cite{winkler_repulsively_2006,folling_direct_2007,petrosyan_quantum_2007,valiente_two-particle_2008,valiente_three-body_2010,preiss_strongly_2015}.
Here we are interested in a non-equilibrium configuration, in the low filling regime, where 
  $M$ particles are initially located on a single site. 
In optical lattices the initialization of the system in one of these states 
is obtained starting from the Mott-Insulator regime and using single-atom addressing
techniques
\cite{weitenberg_single-spin_2011,preiss_strongly_2015,fukuhara_microscopic_2013,fukuhara_quantum_2013,preiss_quantum_2015,islam_measuring_2015}.
The key point here is that, provided that the onsite 
interaction strength $U$ is large enough, the 
resulting state composed of $M>1$ bounded particles on the same site is
stable against dissociation during the time evolution
\cite{winkler_repulsively_2006,petrosyan_quantum_2007,valiente_two-particle_2008,valiente_three-body_2010,lahini_quantum_2012,qin_statistics-dependent_2014},
and  
behaves like an effective single particle. 
  Indeed, as explicitly discussed in \cite{valiente_two-particle_2008,valiente_three-body_2010} 
  for a few values of $M$, the bounded-particle states lie in an energy band which is 
  well-separated (by an energy separation $\propto U$) from other states, provided that 
  $U$ is suitably large. In the following we introduce a general theory to model 
  the effective interactions between stable bounded particles. 

In general when $U_j\gg J_j,\mu_j$  the different Hilbert subspaces $\mathcal H_{M}$ 
spanned by the states with $M$ bounded particles, 
namely $\mathcal H_{M}= \{\ket{\{M\},j}{=}(a_j^\dagger)^M\ket 0/\sqrt{M!} :
j{=}1,\dots,L\}$, are energetically well separated. 
Because of this energy
separation between subspaces, if the initial state is composed of a bound
$M$-particle states, then with a good approximation the dynamics remains
confined inside $\mathcal H_M$.
The resulting effective dynamics can be described with a Hamiltonian  $H^{\rm
eff}_M$ which describes the effective interactions inside $\mathcal H_M$. 
By exploiting the theory presented in \cite{jia_integrated_2015}, which assumes
that the dynamical effective subspace is
energetically separated by rest of the Hilbert space, we find that
\ref{eq:BoseHubbard}, generates in $\mathcal H_M$ the following effective
interaction
\begin{equation}
H_M^{\mathrm{eff}}=
 \left(
 \begin{array}{ccccc}
B^{\mathrm{eff}}_1&J^{\mathrm{eff}}_1&&&  \\
J^{\mathrm{eff}}_{1}& B^{\mathrm{eff}}_{2}&J^{\mathrm{eff}}_{2}&&\\
& \ddots & \ddots&\ddots&\\
&&J^{\mathrm{eff}}_{L{-}2}& B^{\mathrm{eff}}_{L^{-}1}&J^{\mathrm{eff}}_{L{-}1}\\
&&&J^{\mathrm{eff}}_{L{-}1}&B^{\mathrm{eff}}_{L} 
\end{array}
 \right)
\label{eq:HeffM}
\end{equation}
in the basis $\ket{\{M\},j}$, $ j{=}1,\dots,L$. 
The above Hamiltonian describes a quantum walk of a one-dimensional bounded
particle. We mention that, recently, using a different approach, an effective spin-chain model has been obtained to control and manipulate a 1D strongly interacting two specie Bose-Hubbard for quantum communication and computation purposes \cite{volosniev_engineering_2015}.
Quantum walks have been subject to intensive investigations over the 
past years, both with single particles  and multi-particles
\cite{lahini_quantum_2012,qin_statistics-dependent_2014,preiss_strongly_2015,lorenzo_transfer_2015,apollaro_many-qubit_2015,sousa_pretty_2014,paganelli_routing_2013}.
In particular, different schemes have been found to 
engineer the couplings $J^{\rm eff}_j$ and the energies $B^{\rm eff}_j$ such
that the dynamics either produces a perfect state transfer \cite{christandl_perfect_2004,kostak_perfect_2007,kay_perfect_2010}
or a perfect splitting and reconstruction of the initial wave-packet, 
namely a fractional revival
\cite{banchi_perfect_2015,genest_quantum_2015,lemay_novel_2015}. 
From our perspective, a perfect state transfer in the effective subspace would
give rise to a perfect transmission of a bounded particle: namely the state 
$\ket{\{M\},1}\propto (a_1^\dagger)^M\ket 0$ is dynamically transferred to the 
opposite end of the chain 
$\ket{\{M\},L}\propto (a_L^\dagger)^M\ket 0$.
Another important application is the perfect fractional revival, which
effectively generates a beam splitting transformation between the ends 
of the chain. The main reason for its importance is that when bound states are
involved in the perfect splitting transformation, $\ket{\{M\},1}\to
\ket{\{M\},1}+e^{i\phi}\ket{\{M\},L}$, then a $M$-particle NOON state 
$[ (a_1^\dagger)^M +e^{i\phi}(a_L^\dagger)^M]\ket 0$ is produced.

The main idea of our scheme is then to engineer the couplings $J_j$ and the chemical potentials $\mu_j$ in the
Bose-Hubbard, Eq. \eqref{eq:BoseHubbard}, such that
the effective couplings in Eq. \eqref{eq:HeffM} have the suitable pattern for either
state transfer or state splitting (fractional revival). 
The time scale the resulting effective dynamical transformation is approximately given by 
$1/J_{\rm eff}$. However, since the effective hopping in $\mathcal H_M$ involves
$M-1$ ``virtual'' transitions through states which are outside $\mathcal H_M$,
then it is simple to realize that $J^{\rm eff}_j\propto J_j^M/ U_j^{M-1}$, so
$J^{\rm eff}_j$ exponentially decreases with $M$ for large $U$. For larger
$M$-particle bound states the effective evolution thus become slow and the
efficiency of the scheme may be severely affected by environmental effects. 
Because of this, in the next sections we thoroughly analyze the $M=2$ and $M=3$
cases which are more feasible, given  the current experimental capabilities. The overall theoretical scheme is however fully general and can be readily
extended for larger values of $M$. 

\section{Applications}
\subsection{Edge unlocking}

Before focusing on the specific  $M=2$ and $M=3$ cases, we start by 
discussing some general properties of the quantum
walk of bounded particles, to clarify the differences  with the single
particle counterpart. 
We consider the uniform coupling regime, namely $J_j=J$, $U_j=U$,
$\mu_j=\mu$, in the initial state
$\ket{\{M\},1}\propto\left(a_1^\dag\right)^M\ket{0}$. The resulting effective
interaction is 
\begin{equation}
H_M^{\mathrm{eff}}= 
\left(
\frac{J}{U}\right)^{M-1}
\left(
\begin{array}{ccccc}
B^{\mathrm{eff}}_1&J^{\mathrm{eff}}_1&&&  \\
J^{\mathrm{eff}}& B^{\mathrm{eff}}_{2}&J^{\mathrm{eff}}&&\\
& \ddots & \ddots&\ddots&\\
&&J^{\mathrm{eff}}& B^{\mathrm{eff}}_{L^{-}1}&J^{\mathrm{eff}}\\
&&&J^{\mathrm{eff}}&B^{\mathrm{eff}}_{L} 
\end{array}
\right)
\label{eq:HeffM2Unif}
\end{equation}
where $J^{\rm eff}= \mathcal O(J)$, $B_1^{\rm eff}=B_L^{\rm eff}=\mathcal O[J(U/J)^{M-2}]$ while 
$B_j^{\rm eff}$ for $j\neq1,L$ is much smaller than $B_1^{\rm eff}$ (boundary elements may be of
order $\mathcal O(J)$ or less while in the bulk they are even smaller). The
appearance of a larger effective field in the boundaries gives rise to a
phenomenon which is called {\it edge-locking}.  Edge-locked states, which have
been already described in
\cite{haque_self-similar_2010,pinto_edge-localized_2009} for $M\ge3$, can be
understood using the theory of quasi-uniform tridiagonal matrices 
\cite{banchi_spectral_2013}. To describe this phenomenon we consider 
an initial wave-packet localized in site
1 which evolves through the Hamiltonian, Eq. \eqref{eq:HeffM2Unif}, to the wave-packet
$\ket{\psi(t)} = \sum_j (e^{-it H_M^{\rm eff}})_{j1} \ket{\{M\},j}$. Calling 
$H_M^{\rm eff}= VEV^{\dagger}$ the spectral decomposition of the effective
Hamiltonian, then 
$\ket{\psi(t)} = \sum_{kj} e^{-it E_k} V_{1k}V^*_{jk} \ket{\{M\},j}$. Because of the
mirror symmetry and for the properties of quasi-uniform matrices
\cite{banchi_spectral_2013} one finds that $V_{Lk}=V_{1k}(-1)^k\approx 
V_{1k} e^{i L k}$ and $E_k\propto \cos(k)$ where $k$ is the quasi-momentum,
$k=k_j + \mathcal O(L^{-1})$ where $k_j=\pi j/(L+1)$ and $j=1,\dots,L$.
Therefore the quantum walk of the bounded particle displays the standard
expression of a wave-packet evolution \cite{banchi_long_2011}, as
$\langle{\{M\},L}|{\psi(t)}\rangle = \sum_k e^{-i(t E_k-Lk)} |V_{1k}|^2 $ where
$|V_{1k}|^2$ is the probability to excite the quasi-momentum state $k$ by
initializing the system in the first site. 
To simplify the theoretical analysis we assume that $B_j^{\rm eff}\equiv B_{\rm
bulk}^{\rm eff}$ is constant for
$j\neq1,L$ so, without loss of generality, we can set $B_{\rm bulk}^{\rm
eff}=0$. Indeed, the Hamiltonian \eqref{eq:HeffM2Unif}  and 
$H^{\rm eff}_M - B_{\rm bulk}^{\rm eff}\mathbf{1}$ give rise to the same evolution 
aside from an irrelevant global phase. 
Within this description it is now clear that edge-locking appears when $B_1^{\rm
eff}\gg
E_k$, since no quasi-momentum state can be excited by initializing the system in a state where 
the bounded particle is in the first site (namely $|V_{1k}|^2 \approx 0$ for all
the quasi-momentum states). Indeed, in this regime this initialization excites
out-of-band modes  which are localized near the edges and do not propagate. 
As it is clear from Eq. \eqref{eq:HeffM2Unif}, since $B_1^{\rm eff}=B_L^{\rm
eff}=\mathcal O[J(U/J)^{M-2}]$, the edge-locking condition $B_1^{\rm eff}\gg
E_k$ happens when $M\ge 3$, as obtained also in
\cite{pinto_edge-localized_2009}. However, there is another form of
quasi-locking for $M=2$ which is not described in
\cite{pinto_edge-localized_2009}. Indeed, for $M=2$ we find that $B_1^{\rm eff}$ is of the
same order of the energy band $E_k$ of the quasi-momentum states and, as a
consequence, the quasi-momentum states with energy $B_1^{\rm eff}\approx E_k$ 
are the ones involved by the dynamics. Since $E_k\propto\cos(k)$ when $B_1^{\rm
eff}\approx
0$ the relevant excitations consist mostly of quasi-momentum states with  
almost-linear dispersion relation ($E_k\approx k$ around $k=\pi/2$ where
$E_{\pi/2}\simeq 0$). 
These states propagate without dispersion in
the chain and therefore give rise to a high transmission quality. On the other hand, if 
$B_1^{\rm eff}\neq 0$ other states with non-linear dispersion relation are involved, which
drastically lower the transmission quality. 
Because of this, we find that the state
$\ket{\psi}{=}(a^\dagger )^2\ket{0}$ has a long delocalization time from the
initial site during the relevant time $t^*\sim L U/J^2$. 

Because edge localization is detrimental in quantum transfer applications, we
analyze the possibility to ``unlock'' the states by compensating the energy gap
between edge and bulk sites introducing a local static potential both in the
first and the last site of the chain. Edge unlocking can be always obtained for
any value of $M$ by adding suitable local chemical potentials $\mu_j$ around the
edges such that
the effective fields $B_n^{\rm eff}$ are constant over the different sites $n$. 

\subsection{Two Particles}
In this section we analyze the dynamical behavior of a two particle bound
state. We first start from the uniform case, describe the edge unlocking and
then we consider how to engineer the couplings to maximize the transfer of a
bound state and the generation of a NOON state.  

\subsubsection{Edge Unlocking}
\begin{figure}[t]
\centering
\includegraphics[width=1\columnwidth]{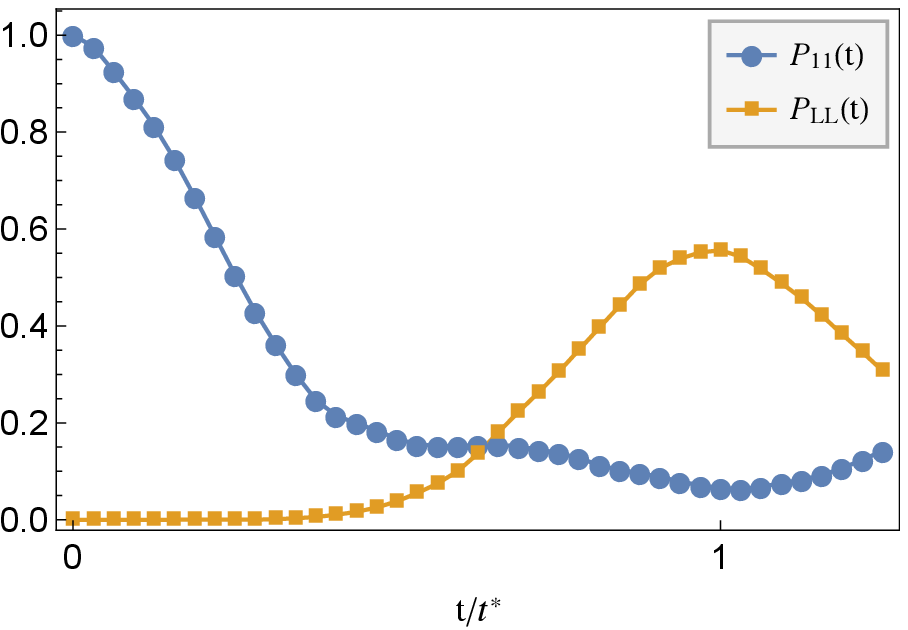}
\includegraphics[width=1\columnwidth]{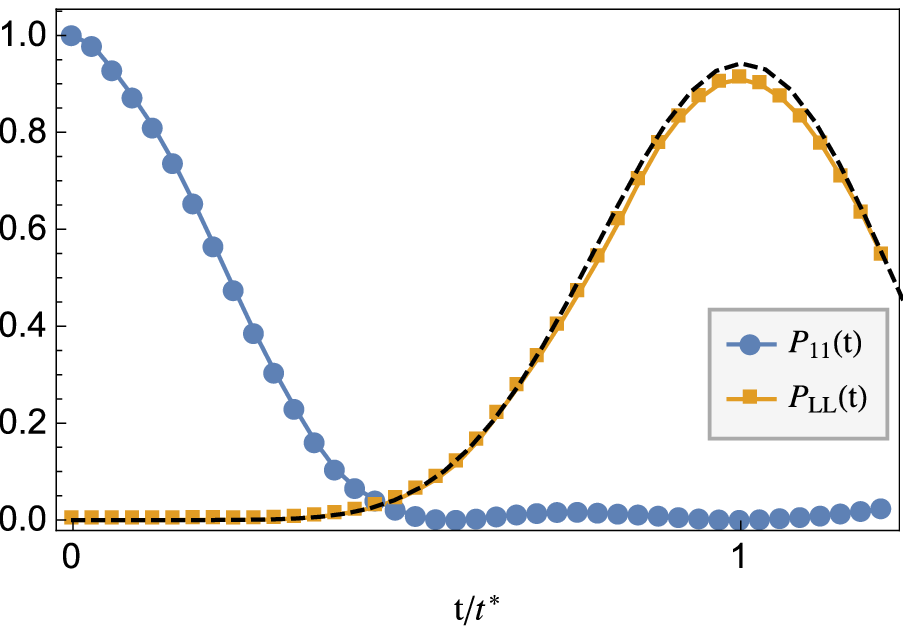}
\caption{Edge delocalization for a two particle bound state: plot of the probability to have a bound particle in the first
    $P_{11}(t)$ and in the last $P_{LL}(t)$ site as a function of the time $t$,
    scaled for the transfer time $t^*\sim L/J_{\textrm{eff}}$ for a uniform chain with $L=5$ and $U/J=5$ in the initial state $\ket{\psi(0)}\propto (a_1^\dag)^2\ket{0}$. We add a local field $\mu_j=-\beta' (\delta_{j,1}+\delta_{j,L})$ with strength 
    $\beta'=0$ (top), and $\beta'=\beta_{\mathrm{opt}}'= J^2/4U$ (bottom). To compare the results with the single particle case we plot, with a dashed black line, the probability $P_{L}(t)=\vert\langle 0\vert a_L \vert \psi(t)\vert^2$ for a single particle initially in $a_1^\dag \ket{0}$ (here $t^*\sim L/J$).}
\label{fig:LocalisationTwoParticles}
\end{figure}

For a uniform chain $J_j=J$, $U_j=U$, $\mu_j=\mu$ we find that the effective
hamiltonian $H_{\mathrm{eff}}$ is the  tridiagonal matrix in Eq. \eqref{eq:HeffM} where 
\begin{equation}
    J^{\rm eff}_j = \frac{J^2}{2U}~, 
\end{equation}
\begin{equation}
    B^{\rm eff}_j = \left\{
    \begin{array}{l@{\quad}cl} 
    \frac{J^2}{2U} + U~, & {\rm for} & j=1,L ~,
        \\
     \frac{J^2}{U} + U~ , & {\rm for} & j\neq 1,L~.
     \end{array}
\right.
\end{equation}
The effect of the inhomogeneities in $B_j^{\rm eff}$ is shown in Fig. 
\ref{fig:LocalisationTwoParticles} (top), where we analyze the dynamics of a
bound particle  initially in $\ket{\psi(0)}\propto\left(a_1^\dag\right)^2\ket{0}$ in a uniform chain with $L=5$ and $U/J=5$. 
We study the probability $P_{ij}(t)$ that after the time $t$ one particle is in
the site $i$ and the other is in the site $j$, namely 
$P_{ij}= \frac{1}{1+\delta_{ij}}|\bra 0 a_i a_j \ket{\psi(t)}|^2$, where
$\ket{\psi(t)}$ is the state evolved for a time $t$. In particular
we plot , as a function of the time $t$, the probability to have the bound particle in the 
first site $P_{11}(t)$  and in the last  site of the chain $P_{LL}(t)$. We
observe that despite the bound particle reaches the last site of the chain at
the transfer time $t^*\sim L / J_{\textrm{eff}}$, the probability to be in the first
site $P_{11}(t^*)$ is still not zero, namely the delocalization time from the
first site is slow compared to the transfer time $t^*$. As described in the
previous section, this is due 
to the difference in effective energies $ B^{\rm eff}_j $ between
the bulk and the edges which favors non-linear excitations which, in turn,
leads to a dispersive dynamics. This difference  between the bulk and the edges
can be made zero by adding two local chemical potentials at the endpoints,
$\mu_j=-\beta'\left(\delta_{j,1}+\delta_{j,L}\right) $, where $\beta'=J^2/4U$. As it can be seen in Fig. \ref{fig:LocalisationTwoParticles} (bottom), when the $\beta'$ field is added, the delocalization time from the first time $P_{11}(t^*)$ (and consequently the transfer fidelity $P_{LL}(t^*)$) is strongly increased. We compare the results obtained for the transfer of a bound state with the propagation of a single particle in the lattice, initially in $a_1^\dag \ket{0}$, by plotting in Fig. \ref{fig:LocalisationTwoParticles} (bottom) the probability $P_{L}(t)=\vert\langle 0\vert a_L \vert \psi(t)\vert^2$ (single particle data are scaled for their transfer time $t^*\sim L/J$). The difference between the single particle and the bound particle results depends on the finite value of the interaction chosen ($U/J=5$ in Fig. \ref{fig:LocalisationTwoParticles}). Indeed the agreements with a single particle behavior is very high as long as $U/J$ is large enough. 

Because the effective model in Eq. \eqref{eq:HeffM} is valid in the 
regime $U/J\gg1$, we analyze deviations from the theoretical value of $\beta'$,
by evaluating the dynamics with exact diagonalization techniques as in
\cite{compagno_toolbox_2015,banchi_perfect_2015}. Once initialized the system in
$\ket{\psi(0)}\propto(a_1^\dag)^2\ket{0}$ we numerically find the value of
$\beta'$ that maximize the probability to find, at the transfer time $t^*\sim
L/J_{\mathrm{eff}}$, the bound particle in the last site of the chain
$P_{LL}(t^*)$. 
We numerically find that, for a two particle bound state, the optimal values of $\beta'$ completely agree with the theoretical model  $\beta'= J^2/4U$ independently from the length of the chain, as long as  $U/J\gtrsim 5$. 

When the hopping term $J$ and the onsite interaction $U$ have comparable
amplitude, both the bound states and the single particle states contribute to
the dynamics \cite{preiss_strongly_2015}. We expect that by increasing the
onsite interaction the effects of the free particle states are reduced while the state
transfer fidelity of a bound particle should converge to a constant value. 

By analyzing $P_{LL}(t^*)$ we find that, when the optimal 
value for the localized field $\beta'=J^2/4U$ is added in a uniform chain,
 values of the onsite interaction above $U/J\gtrsim 4$ guarantee an almost
constant value of transfer fidelity for a two particle bound state. 
\subsubsection{Optimal State Transfer of a Two Particle Bound state}\label{sec:OptimalTransfer2Particles}
\begin{figure}[t]
\centering
\includegraphics[width=1\columnwidth]{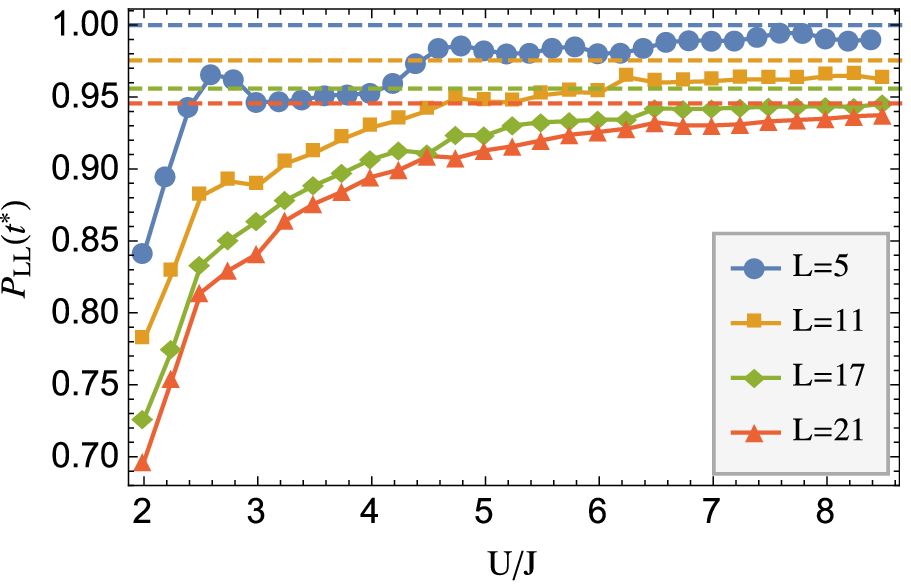}
\includegraphics[width=1\columnwidth]{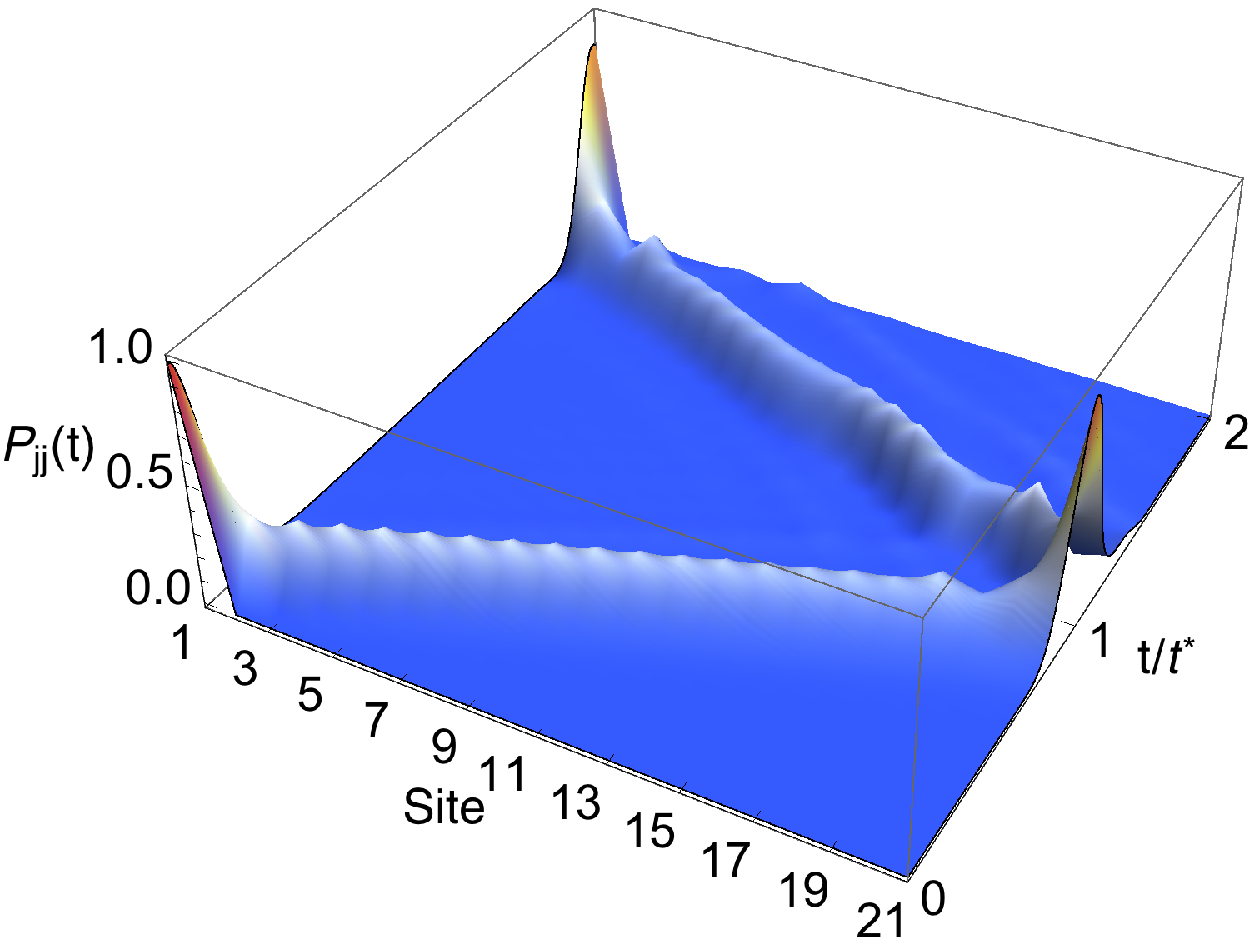}
\caption{Optimal transfer of a two particle bound state: (top) analysis of the transfer fidelity $P_{LL}(t^*)$ for the initial state
$\ket{\psi(0)}\propto(a_1^\dag)^2\ket{0}$ as a function of the onsite interaction
$U/J$ when the optimal transfer scheme $J_1=J_L=J_0$ and $J_j=J$ for $j\neq
(1,L)$ is included in the model \ref{eq:BoseHubbard}. Here we also add two
local impurities $\mu_j= -\beta_1 (\delta_{j,1}+\delta_{j,L})$ and $\mu_j= -\beta_2 (\delta_{j,2}+\delta_{j,L-1})$ where $\beta_1=\left(J_0^2 - 2 J^2\right)/2 U$ and $\beta_2=\left(J_0^2 -  J^2\right)/2 U$ to eliminate the edge-locking effect. The $J_0$ value is chosen by numerically maximizing the transfer fidelity in a single particle manifold \cite{banchi_long_2011}. 
To compare the difference between a single particle and a bound state, we plot (with a dashed 
line) also the single-particle transfer fidelity 
$P_{L}(t^{*})=\vert \langle 0\vert a_L\vert\psi(t^*)\rangle\vert^2$ obtained for a system initially in $a_1^\dag \ket{0}$. (bottom) Probability $P_{jj}(t)$ to have a two particle bound state in site $j$ at time $t/t^*$ for a minimal engineered chain with $L=21$ and $U/J=8$.}
\label{fig:MaxFidelityOptimalM2}
\end{figure}

The state transfer efficiency of a two particle bound state  in a uniform chain
can be improved by suitably tuning the tunneling couplings in the model in Eq. \eqref{eq:BoseHubbard}. Because a full
engineering could be too demanding, here we consider the effect of a minimal
engineering scheme \cite{banchi_long_2011,apollaro_99$$-fidelity_2012}, which consists of tuning the first and the last tunneling
terms to $J_1=J_{L-1}=J_0$ while  the rest of the chain has uniform couplings 
$J_j=J$. In this case the effective Hamiltonian
\eqref{eq:HeffM} has effective interactions 
\begin{equation}
    J^{\rm eff}_j =\left\{ \begin{array}{l@{\quad}cl}
   \frac{J_0^2}{2U} & {\rm for} & j=1,L-1~, 
    \\
    \frac{J^2}{2U} & {\rm for} & j\neq 1,L-1~,  
    \end{array}
    \right.
\end{equation}

\begin{equation}
B^{\rm eff}_j = \left\{  \begin{array}{l@{\quad}cl} 
 \frac{J_0^2}{2U} + U & \mathrm{for} & j=1,L,  \\  
\frac{J_0^2}{2U} + \frac{J^2}{2U}  + U &  \mathrm{for} & j=2,L-1, \\
  \frac{J^2}{U} + U& {\rm for} & j= 3,\dots,L-2.
\end{array}
\right.
\end{equation}
To maximize the transfer fidelity one has then to remove the difference between 
the effective energies $B^{\rm eff}_j$, and to optimize the values of $ J^{\rm
eff}_j $ to achieve an optimal ballistic dynamics. Since the dynamics occurs in 
the effective subspace one can use the analytical theory presented  in \cite{banchi_long_2011} 
to find the optimal value of  $J_0^{\rm eff} \equiv \frac{J_0^2}{2U} $, given
that the rest of the sites are coupled with a hopping strength 
$J^{\rm eff} \equiv \frac{J^2}{2U} $. Given the simple relationship between 
$J_0^{\rm eff}$ and the strength $J_0$ of the Bose-Hubbard tunneling between the
edges and the bulk, it is then straightforward to obtain $J_0$. Once $J_0$ is
found, 
one needs to add local chemical potentials in both the first and the last two sites to remove
the local energy difference in $B^{\rm eff}_j$. By using our effective
Hamiltonian expansion,
we find that the state transfer is
maximized by introducing two pairs of local fields, respectively $\mu_j=
-\beta_1 (\delta_{j,1}+\delta_{j,L})$ and $\mu_j= -\beta_2
(\delta_{j,2}+\delta_{j,L-1})$, with strengths $\beta_1=\left(J_0^2 - 2
J^2\right)/2 U$ and $\beta_2=\left(J_0^2 -  J^2\right)/2 U$. In Fig. 
\ref{fig:MaxFidelityOptimalM2} (top) we show the results obtained for the transfer
fidelity $P_{LL}(t^*)$ as a function of $U/J$ when we use minimal engineering
and the compensating fields $\beta_1$ and $\beta_2$. We observe a significant
improvement of transfer fidelity compared to the results for a uniform chain.
In Fig. \ref{fig:MaxFidelityOptimalM2} (bottom) we also highlight the difference between the single-particle dynamics 
and the bound-particle case for finite interaction $U$. 
As expected, for strong
inter-particle interaction $U$, a bound state behaves as a single particle
state. To highlight that minimal engineered schemes already have a significant impact in reducing the dispersion in the system, in Fig. \ref{fig:MaxFidelityOptimalM2} (bottom) we show the dynamics of a bound state in a lattice. Specifically, we plot the probability $P_{jj}(t)=\vert \langle 0\vert a_j^2 \vert \psi(t)\rangle\vert^2/2$ to have a two particle bound state in site $j$ as a function of the time $t/t^*$, for a minimal engineered chain with $L=21$ and $U/J=8$.

In analog fashion the Bose-Hubbard Hamiltonian couplings can be tuned so that the
effective Hamiltonian coincides with the one allowing perfect state transfer
\cite{christandl_perfect_2004}, though this is much more demanding because it
requires the engineering of all tunneling rates $J_j$ and all the chemical
potentials $\mu_j$.

\subsubsection{Environmental effects}

State transfer schemes are generally robust against static imperfections in the
couplings \cite{zwick_robustness_2011,pavlis_evaluation_2016}. On the other
hand, we explicitly test the robustness of our scheme against dynamical environmental effect, specifically dephasing due to spontaneous emission, which represents the main source of decoherence in optical lattices. The dynamics of the system in the lowest band is typically modeled as a Master equation in Lindblad form \cite{kordas_non-equilibrium_2015,pichler_nonequilibrium_2010,sarkar_light_2014}:
\begin{equation}
\dot{\rho}=-i\left[H_{BH},\rho\right]+\Gamma \sum\limits_i  \left(n_i \rho n_i - \frac{1}{2} n_i n_i \rho - \frac{1}{2} \rho n_i n_i\right).
\label{eq:MasterEqDecoherence}
\end{equation}
Here $\Gamma$ is the effective scattering rate, and $H_{\mathrm{BS}}$ is the Bose-Hubbard Hamiltonian \ref{eq:BoseHubbard}. We numerically solve Eq.  \eqref{eq:MasterEqDecoherence} as shown in detail in \ref{sec:Decoherence}. 

\begin{figure}[t]  
\includegraphics[width=1\columnwidth]{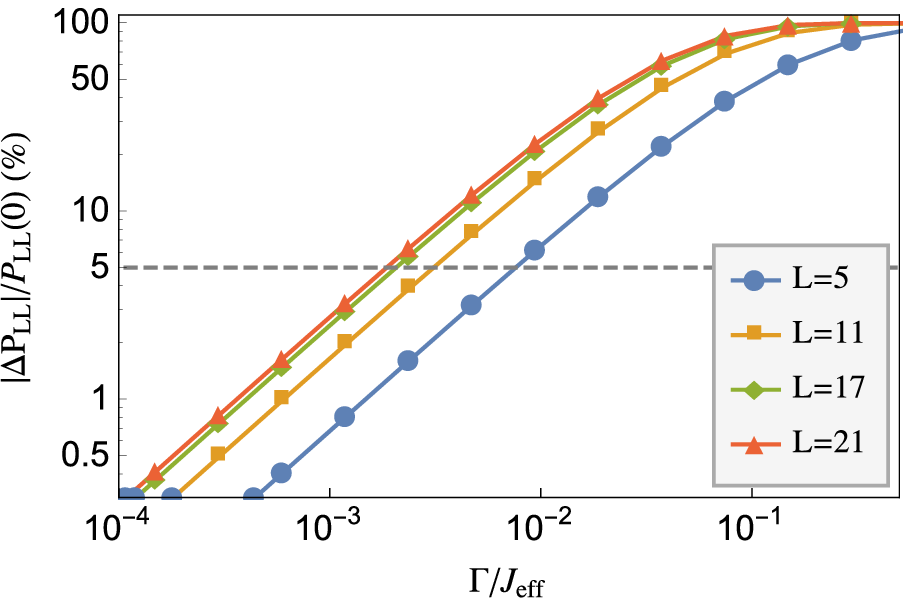}
\includegraphics[width=1\columnwidth]{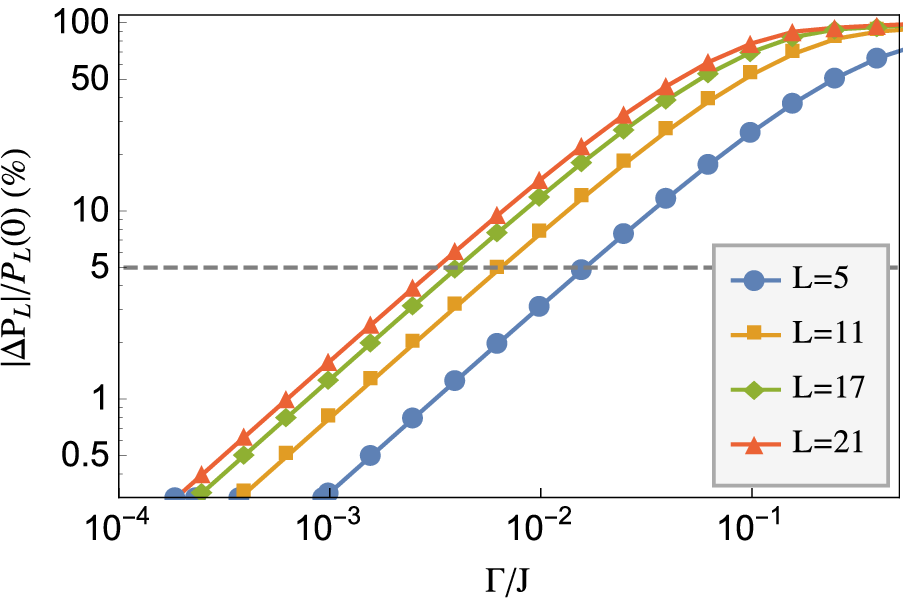}
\caption{Decoherence effects for a two particle bound state: (top) relative variation $\Delta P_{LL}/P_{LL}(t^*,\Gamma=0)$ of the transfer fidelity $P_{LL}(t^*)$ with respect to the case in absence of decoherence, for a uniform chain with $U/J=3$ and length $L$. Here $\Delta P_{LL}=\vert P_{LL}(t^*,\Gamma)-P_{LL}(t^*,\Gamma=0)\vert$  and the dashed grey line is a threshold of a relative variation of the $5\%$. Several chain length $L$ are considered. 
(bottom) Relative variation $\Delta P_{L}/P_{L}(t^*,\Gamma=0)$ for a single particle state initially in $a_1^\dag\ket{0}$.
}
\label{fig:MaxFidelityBPrimeGammaU3}
\end{figure}

No relevant edge field optimal strength $\beta'$ deviations are found when decoherence effects are introduced, for $\Gamma/J_{\mathrm{eff}}<0.1$, where $J_{\mathrm{eff}}=J^2/2U$. In Fig. \ref{fig:MaxFidelityBPrimeGammaU3} (top) we show how the transfer fidelity $P_{LL}(t^*)$ is affected as a function of the damping rate $\Gamma/J_{\mathrm{eff}}$ in Eq. \eqref{eq:MasterEqDecoherence} for $U/J=3$.
To better evaluate the difference with the zero decoherence case, we show in Fig.  \ref{fig:MaxFidelityBPrimeGammaU3} the relative variation $\vert \Delta P_{LL}(t^*)\vert/P_{LL}(0)=\vert P_{LL}(\Gamma)-P_{LL}(\Gamma=0)\vert/P_{LL}(\Gamma=0) $ with respect to the no decoherence case, as a function of the damping parameter $\Gamma/J_{\mathrm{eff}}$. We observe deviations of less than the $5\%$ for $\Gamma/J_{\mathrm{eff}}\simeq 10^{-2}- 10^{-3}$ for chain lengths between $L\in\left\{5,\ldots,21\right\}$ which are typical values for blue detuned optical lattices \cite{sarkar_light_2014,pichler_nonequilibrium_2010}. In Fig. \ref{fig:MaxFidelityBPrimeGammaU3} (bottom) we show the effects of the decoherence in the state transfer fidelity for a single particle state, initially in $a_1^\dag\ket{0}$.

\subsubsection{NOON State Generation with a Two Particle bound state}
In this section we consider an imperfect fractional revival by considering a
uniform evolution, though the present results can be extended with a further
engineering to achieve perfect fractional revivals. 

\begin{figure}[t]
\centering
\includegraphics[width=1\columnwidth]{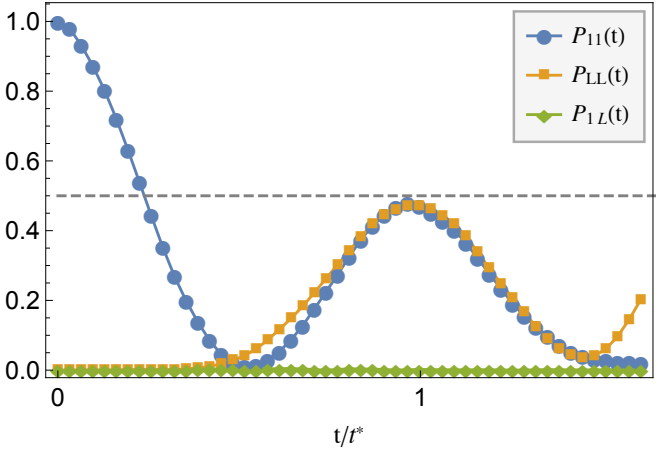}
\caption{Two particle NOON state: plot of the probabilities $P_{ij}(t)$ to have one particle in site $i$ and the other in $j$, as a function of the time, in unit of the transfer time $t^*$, for two particles initially in $\ket{\psi(0)}\propto\left(a_1^\dag\right)\ket{0}$. Here we consider a uniform chain with an impurity $\mu_j=-\beta \delta_{j,L/2+1}$ where $\beta=0.789 (J^2/2U)$ in a a chain with $U/J=5$ and length $L=5$. The absence of the $P_{1L}(t^*)$ term is an evidence that the output state at $t=t^*\simeq U L/J^2$ is the NOON state with two particles. The grey dashed line represents the results for an ideal lossless NOON state generation. 
}
\label{fig:P11SuppressionM2}
\end{figure}

We consider the simplest
scheme where the wave-packet splitting is achieved by using a local barrier in
the middle of the chain, as shown in Fig.  \ref{fig:BPUScheme} and discussed in 
\cite{compagno_toolbox_2015}.
We set the value $\beta'=J^2/4U$ for the edge fields $\mu_j=-\beta'(\delta_{j,1}+\delta_{j,L})$ 
to remove edge locking. Then we add a local field in the middle of the chain
$\mu_j=-\beta \delta_{j,L/2+1}$ to trigger a wave-packet splitting
\cite{compagno_toolbox_2015}. 
Indeed, the extra barrier favors the splitting of the propagating bound
particle wave-packet into a transmitted and reflected component. It has been
shown in 
\cite{compagno_toolbox_2015} that for single particle quantum walk, the optimal
50/50 splitting is obtained when the strength $\beta$ of this extra barrier is 
equal to the hopping rate. 
We expect that when the  bound particle impinges
the splitting field because it behaves like an effective single particle, the
output state, measured at the endpoints, will be
$\ket{\psi(t^*)}_{1L}=\frac{1}{\sqrt{2}}\left(\ket{2,0}+i \ket{0,2}\right)$, namely
we generate a NOON state with two particles (here $\ket 2 =
(a^\dagger)^2\ket0/\sqrt 2$). On the other hand if the two particles are non
interacting $U/J=0$, the effect of the splitting field is to produce also a
non-zero probability $P_{1L}(t^*)$ to have one particle in each end
\cite{laloe_quantum_2011}. We show, for a $L=5$ chain with $U/J=5$ in Fig. 
\ref{fig:P11SuppressionM2} that for a bound particle that term is suppressed at
the transfer time $t^*$, as expected from a bound particle effective evolution.
Therefore we can conclude that the output state is, when $U$ is large enough, the
NOON state with two particles, apart from a damping factor due to dispersion. 
By
performing an effective Hamiltonian expansion in the onsite interaction term we
show that to have a balanced splitting in the effective space, the strength of
the splitting field in the real chain must be $\beta=\beta^{50/50}=J^2/2U$, in
the limit $L\gg1$. 

\begin{figure}[t]
\centering
\includegraphics[width=1\columnwidth]{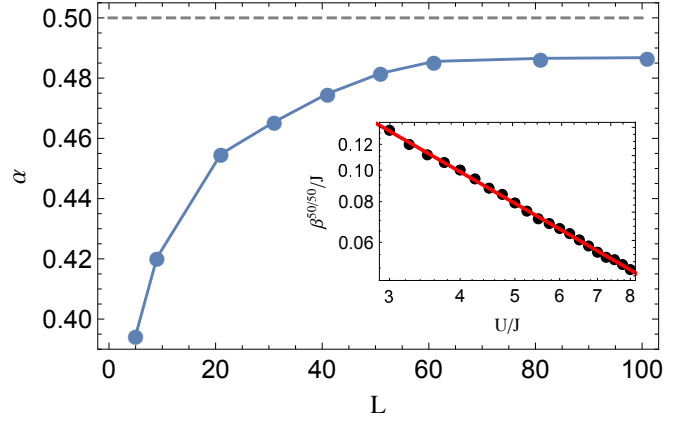}
\caption{Finite-size effects in the two particle NOON state generation: analysis of the scaling factor $\alpha$ where $\beta^{50/50}=\alpha
J^2/U$, as a function of the chain length $L$ for generating a two particle NOON
state. The grey line represents the theoretical results from the effective
Hamiltonian theory, which holds for $L\gg 1$. (Inset) Analysis of the optimal value of $\beta$ of the local field $\mu_j=-\beta \delta_{j,L/2+1}$ which produces the NOON state with two particles as a function of $U/J$ for $L=5$. The red line is the fit $\beta=\beta^{50/50}=0.395J^2/U$.}
\label{fig:BSM2UniformFiniteFactor}
\end{figure}

Finite length corrections  are found numerically by finding the $\beta$ value
for which the difference $P_{11}(t^*)-P_{LL}(t^*)$ is zero. As shown in the inset in Fig. 
\ref{fig:BSM2UniformFiniteFactor}, for a $L=5$ uniform chain, $\beta^{50/50}$ scales as
$1/U$. The finite length factors, found from a fit over the data for several chain lengths, are shown in Fig.  \ref{fig:BSM2UniformFiniteFactor}. By
increasing the chain length $L$ the $\beta^{50/50}$ values are closer to
$0.5 J^2/U$ in agreement with the effective Hamiltonian analysis. 
We underline that the NOON state creation efficiency can be made arbitrarily
close to 100\% by tuning the couplings in the effective bound particle subspace
using the techniques for perfect splitting developed in
\cite{banchi_perfect_2015,genest_quantum_2015}  which require a complete engineering of the couplings in the Hamiltonian \eqref{eq:BoseHubbard}. 
In Section \ref{sec:NOONMeasurement} we propose two methods for detecting the NOON state generated by measuring interference fringes. 

\subsubsection{Even Chains}
Here we clarify that our scheme is not limited to odd length chains but also can
be applied to even chains by tuning both the middle tunneling coupling strength
$J_{L/2}$, and adding two pairs of local fields, respectively
$\mu_j=-\beta_1(\delta_{j,1}+\delta_{j,L})$ and
$\mu_j=-\beta_2(\delta_{j,2}+\delta_{j,L-1})$ in the Hamiltonian \eqref{eq:BoseHubbard}. Using the results in
\cite{compagno_toolbox_2015} for the splitting of a single particle we find,
from the effective Hamiltonian model, that the optimal coupling strengths to generate a two particle NOON state between the endpoints of a uniform chain are respectively $J_{L/2}=J(\sqrt{2}-1)^{1/2}$, $\beta_1=J/4U$ and  $\beta_2=J(2-\sqrt{2})/4U$. 

\subsection{Three Particles }
The extension of the previous scheme to more than two particle bound states enhances its non-classical state generation capabilities, namely towards realizing small cat states between remote sites. 
As before the onsite interaction generates bound states with three particles when initially located in the same site. Similarly to the two particle case, one would expect that the results of the splitting process, when the onsite interaction is strong enough, is to produce a NOON state with $N=3$. 

As for the two-particle case, for large onsite interactions the effective
evolution in the bound particle subspace is described by Eq. \eqref{eq:HeffM}. 
We consider a uniform chain, while a minimally engineered model is discussed in 
 \ref{app:j0}. For a
uniform chain the effective hopping is 
$J_{\mathrm{eff}}=3J^3/16 U^2$. Moreover, to remove edge locking and compensate the energy 
gap between the endpoint sites and the bulk of the chain we have to introduce 
two local fields 
$\mu_j=-\beta'\left( \delta_{j,1}+\delta_{j,L} \right)$. From our expansion we find for a
uniform chain that $\beta'=J^2/8U$. To check for finite-size correction to the
above analytical prediction 
we numerically analyze the value of $\beta'$ for several chain length $L$ for
the initial state $\ket{\psi(0)}\propto\left(a_1^\dag\right)^{3}\ket{0}$ as a
function of the onsite interaction $U$. We find that with high accuracy the estimated field $\beta'= J^2/8U$ is independent on $L$.
We analyze the probability $P_{LLL}(t^*)$ to have three particles in the site $L$ after time $t^*$ for a uniform chain as a function of the onsite interaction. We find that values of the onsite interaction above $U/J\gtrsim 4$ guarantee an almost constant value of transfer fidelity for chain lengths $L\in \left\{5,\ldots,21\right\}$.

\begin{figure}[t]
\centering
\includegraphics[width=1\columnwidth]{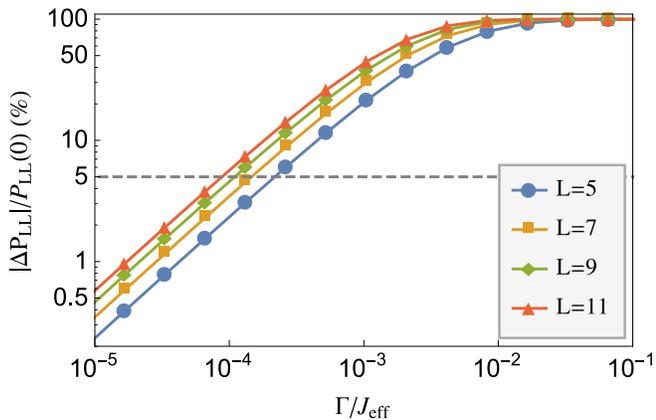}
\caption{Decoherence effects for a three particle bound state: relative variation $\Delta P_{LL}/P_{LL}(t^*,\Gamma=0)$, with respect to decoherence free case, for a uniform chain with $U/J=2$. Here $\Delta P_{LL}=\vert P_{LL}(t^*,\Gamma)-P_{LL}(t^*,\Gamma=0)\vert$  and the dashed grey line is a threshold of a relative variation of the $5\%$. Several chain length $L$ are considered.}
\label{fig:DecoM3}
\end{figure}

In Fig.  \ref{fig:DecoM3} we show the effect of decoherence due to spontaneous emission, Eq. \eqref{eq:MasterEqDecoherence} for several chain lengths $L$ as a function of the decoherence rate $\Gamma/J_{\mathrm{eff}}$ where $J_{\mathrm{eff}}=3 J^3/16 U^2$. We observe relative variation of less than the $5\%$ with respect to the decoherence free case, for $\Gamma/J_{\mathrm{eff}}\lesssim 1.3\times 10^{-4}$ up to $L=7$ sites. 
The state transmission fidelity of a three bound particle state can be optimized by engineering the end tunneling couplings of the chain, as shown in \ref{app:j0}. In this case to bypass the edge-localization effects we also need to add a local chemical potential tuning in the first two and last two sites of the chain. 
\subsubsection{NOON State generation for a three particle bound state}
\begin{figure}[t]
\centering
\includegraphics[width=1\columnwidth]{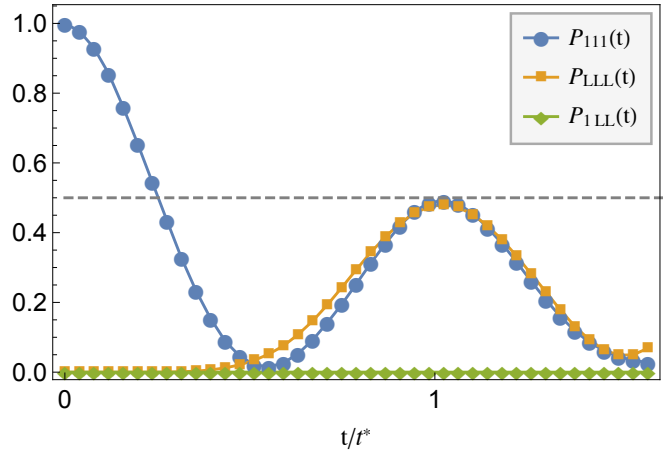}
\caption{Three particle NOON state: joint probabilities $P_{ijk}(t)$ as a function of the time, in unit of
the transfer time $t^*$, for three particles initially in
$\ket{\psi(0)}\propto\left(a_1^\dag\right)\ket{0}$ for a uniform chain with an impurity
$\beta=0.099J^3/U^2$ and $\beta'=J^2/8U$in the middle of the chain with
$U/J=5$ and length $L=5$. The absence of the $P_{1LL}(t)$ term is an evidence
that the output state at $t=t^*$ is the NOON state with two particles. The grey
dashed line represents the results for an ideal lossless NOON state generation.
We found also that $P_{1LL}(t)=P_{11L}(t)$. }
\label{fig:P11SuppressionM3}
\end{figure}

\begin{figure}[t]
\centering
\includegraphics[width=1\columnwidth]{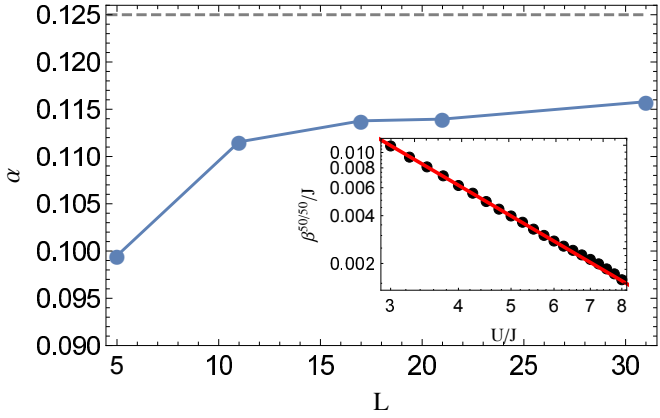}
\caption{Finite-size effects in the three particle NOON state generation: analysis of the scaling factor $\alpha$ as a function of the chain
    length $L$ for a three particle bound state, where $\beta^{50/50}=\alpha
J^3/U^2$. The dashed grey line represents the theoretical value from the
effective Hamiltonian description. (Inset) Analysis of the optimal value of $\beta$ to produce the NOON state with three particles as a function of $U/J$ in a uniform chain with length $L=5$. The red line is the fit $\beta^{50/50}=\alpha J^3/U^2$ where $\alpha=0.099$.}
\label{fig:BSM3UniformFiniteFactor}
\end{figure}

In the non-interacting case $U/J=0$ when three particles are initially located
in the first site $\ket{\psi(0)}\propto(a_1^\dag)^3\ket{0} $ 
an ideal beam splitter transformation generates as output a state with
probabilities \cite{laloe_quantum_2011}: $P_{111}=P_{LLL}=1/8,\
P_{1LL}=P_{11L}=3/8$ where we define $P_{jkl}(t)=\frac{\vert \langle 0 \vert a_i a_j
a_k \vert \psi(t)\rangle\vert^2}{1 + \delta_{ij} + \delta_{jk} + \delta_{ik} +
2\delta_{ij}\delta_{jk}}$ the probability to have the three particles in  the
sites 
$i,j,k$. We expect that when the onsite interaction is strong enough, the bound
particle behaves as an effective single bound particle, thus the terms
$P_{1LL},P_{11L}$ are suppressed and the output state at the end points
effectively results in the NOON state
$\ket{\psi(t^*)}_{1L}=\frac{1}{\sqrt{2}}\left(\ket{3,0}+i \ket{0,3}\right)$, where 
$\ket 3 =(a^\dagger)^3\ket0/\sqrt{6}$. In Fig. \ref{fig:P11SuppressionM3} we
plot, as a function of time (in $t^*\simeq L/J_{\mathrm{eff}}$ units) the
probability to have a three particle bound state respectively, in the first site
$P_{111}(t)$, in the last $P_{LLL}(t)$ and one particle in the first site and
two in the last $P_{1LL}(t)$ in a uniform chain with $U/J=5$ and $L=5$. Here we
set the edge fields strength to $\beta'=J^2/8U$ and we find numerically that the
splitting field to have a balanced splitting is
$\beta=\beta^{50/50}=0.099J^3/U^2$. The absence of the term $P_{1LL}(t^*)=P_{11L}(t^*)$ is an evidence that a NOON state with three particle is generated between the edges of the chain. 
From the effective Hamiltonian description we find
that to generate a balanced splitting of a bound three particle  wave-packet, we
need to add a local field $\mu_j=-\beta^{50/50}\delta_{j,L/2+1}$ whose
strength, when $L\gg1$, is $\beta^{50/50}=J^3/8U^2$ (as explained in 
appendix $B$). 
However, finite size corrections change the value of  $\beta^{50/50}$, and 
by performing a numerical fit over the data for a uniform chain with $L=5$
(whose results are shown in the inset of Fig. \ref{fig:BSM3UniformFiniteFactor}), we find that $\beta^{50/50}$ scales with the onsite interaction as $\beta^{50/50}=\alpha J^3/U^2$ where $\alpha\simeq 0.099$.
Deviations from the theoretical value of $\beta^{50/50}=J^3/8U^2$, which are shown in Fig. \ref{fig:BSM3UniformFiniteFactor}, have been found by analyzing the results obtained for several chain length $L$. Here the dashed grey line represents the theoretical value of the coefficient $\alpha$ of the splitting field for $L\gg 1$. 

\section{NOON State Verification} \label{sec:NOONMeasurement}
The creation and the detection of a NOON state can be revealed by measuring the
interference fringes in a Mach-Zehnder setup. 
After initializing the bound particles in the initial state 
$\ket{\psi(0)}\propto\left(a_1^\dag\right)^N\ket{0}$, the NOON state is
generated by the splitting field in the middle of the chain, as previously
discussed. By freezing the dynamics of the system at the transfer time $t^*$, a controllable phase factor can be added using a local field in the last site of the chain. Finally, once lowered the lattice potential, a second beam splitter operation is performed by the splitting field which produces interference fringes at the endpoints of the chain at $2 t^*$. 

For an ideal lossless transformation the state at the two boundary sites of the
chain at the transfer time $t^*$ would be 
\begin{equation}
    \ket{\psi(t^*)}_{1L}=\frac{1}{\sqrt{2}}\left(\ket{N0}+i \ket{0N}\right)~.
\end{equation}
Once we apply the phase transformation $\Phi=\mathrm{diag}(1,e^{i\phi})$ (namely a
phase shift on site $L$), a second ideal beam splitting transformation would produce 
the output state at time $2t^*$:
\begin{equation}
    \ket{\psi(2t^*)}_{1L}=\frac{1}{2}\left[\left(1-e^{i N\phi}\right)\ket{N0}+i
\left(1+e^{i N\phi}\right)\ket{0N}\right]~,
\end{equation}
where the phase accumulated is $N\phi$ with $N$ being the number of particle in
the NOON state. Therefore the presence of a NOON state is revealed by measuring
the interference fringes (i.e. the probability to have $N$ particle in the first site as a function of $\phi$). 
Although previous experimental results measured just the parity of single sites
(which would exclude a direct observation of the $N=2$ case discussed so far), 
this detection issue in optical lattice has been recently circumvented up to
four particles in the same site \cite{zupancic_ultra-precise_2016}. 

We evaluate numerically the interference fringes for a two particle bound state
in a uniform chain with length $L=5$ and $U/J=5$. 
To introduce a controllable phase factor between the endpoints of the chain we
freeze the dynamics at time $t^*\simeq LU/J^2$, by increasing the lattice
potential depth, then we apply a local field in the last site.  
The Hamiltonian \eqref{eq:BoseHubbard} is then quenched at $t^*$ to 
\begin{equation}
H'= \sum\limits_{j=1}^{L}U n_j (n_j-1)-\beta_L n_L\ .
\end{equation}
We let the system evolve for a time $t'$ and the phase difference generated
between site $L$ and $1$ is $\phi =\beta_L t'$. Then for $t>t'$ the lattice
potential is lowered again and the dynamics is described again by the
Hamiltonian \eqref{eq:BoseHubbard}. Finally we let the system evolve and we
evaluate the probability $P_{11}$ to have the bound particle in the first site
at the transfer time. An alternative approach, discussed in
\cite{compagno_toolbox_2015} for the single particle case, is to add a further
step-like potential on the right-half of the effective chain, which corresponds
to a piecewise constant potential in the Bose-Hubbard model. 

\begin{figure}[t]
\centering
\includegraphics[width=1\columnwidth]{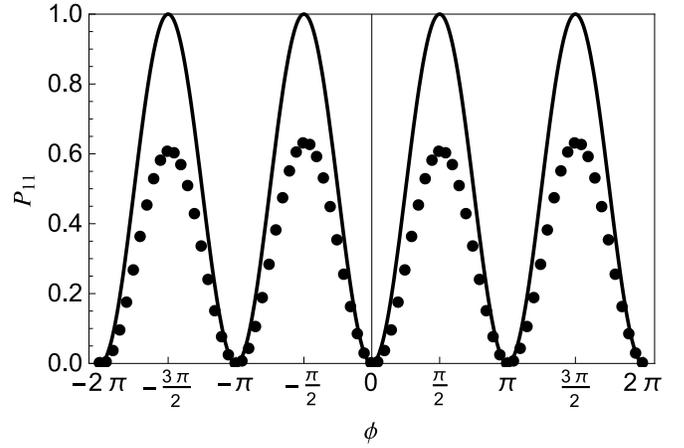}
\caption{NOON state detection with bound particles: interference fringes in a Mach-Zehnder scheme for a two particle
    bound state for a chain with $L=5$ and $U/J=5$. We plot the probability
    $P_{11}(t^*)$ to have the bound state particle in the first site after an time evolution for $t^*\simeq L U/J^2$, when a phase factor $\phi$ is introduced in the system. The line data represent the result for an ideal lossless Mach-Zehnder transformation.}
\label{fig:PlotFringesBPUL5U5}
\end{figure}

In Fig. \ref{fig:PlotFringesBPUL5U5} we show the results for $P_{11}$ as a
function of the phase factor $\phi$. By comparing our data with the results of
an ideal lossless transformation (line in Fig. \ref{fig:PlotFringesBPUL5U5}) we
observe the interference fringes are in the same positions as in the ideal
transformation. The influence of the chain dispersion reduces the height of the
peaks compared to the ideal case. However the efficiency of this scheme can be
pushed up to 100\% by engineering the chain couplings
\cite{compagno_toolbox_2015,banchi_perfect_2015} in the effective subspace of
bound particles.
This method can be easily extended to bound states with a higher number of
particles. 

\begin{figure}[t]
\centering
\includegraphics[width=1\columnwidth]{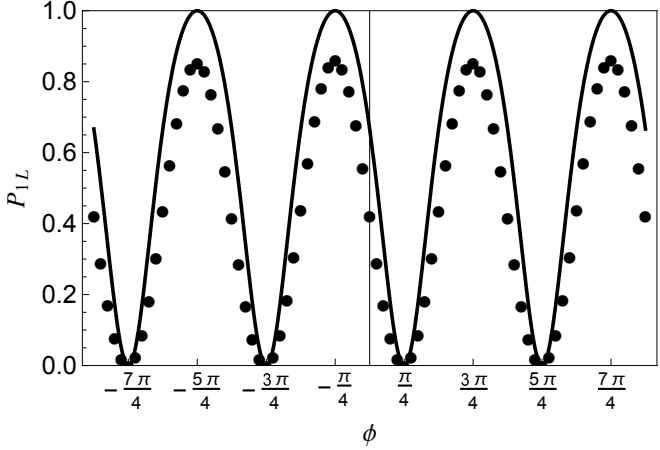}
\caption{NOON state detection after quenched inter-particle interaction: interference fringes after the quench to $U/J=0$. The chain has length $L=5$ and we set $U/J=5$ for generating the two particles NOON state in the edges at time $t^*\simeq L U/J^2$. Once the NOON state is generated the dynamics is frozen by increasing the lattice depth, then a controllable phase factor $\phi$ is added by tilting the lattice. Once the inter-particle interaction is quenched to $U/J=0$ we let the system evolve and we measure the probability $P_{1L}(t'')$ where $t''\simeq L/J$ is the transfer time of the free chain. }
\label{PlotFringes11QuenchedL5U5}
\end{figure}

An alternative approach to detect the NOON state for $N=2$ is to quench the
inter-particle interaction ($U/J=0$, i.e. via Feshbach resonances) just after
the phase factor is added in the system. In the lossless case the final state of
the two boundary sites is  
	\begin{equation*}
        \ket{\psi(t'')}_{1L} \propto\left[(1-i e^{iN\phi})\ket{20}+(i
        e^{iN\phi}-1)\ket{02}+2 i (1+e^{iN\phi})\ket{11} \right] ~.
	\end{equation*}
Here $t''$ is the transfer time of free particles in the lattice $t''\simeq L/J$. The probability to find one particle in each end at $t''$ is
\begin{equation}
P_{1L}(t'')=\frac{2\left(\sin N\phi-1\right)}{\sin N\phi -3}~.
\label{eq:IdealFreeMZ}
\end{equation}
From the latter we see with a choice of $\phi=-5\pi/4$ the output state results
in $\ket{\psi(t'')}_{1L}=\ket{11}$, which can be measured using single particle
fluorescence techniques. The latter scheme has two main advantages: first of all
it circumvents the parity projection measurement issue, because the fringes
measurement require only single atom detection. In second place the decoherence
influence is reduced, because after the phase factor is added it exploits free
particle propagation, which is faster compared to the bound state case. In Fig. \ref{PlotFringes11QuenchedL5U5} we show the results obtained for the probability to observe one particle in each end, in a chain with $L=5$ and $U/J=5$, at time $t''$, compared to the lossless case in Eq. \eqref{eq:IdealFreeMZ}. 
\section{Quantum enhanced metrology}
As shown in Fig. \ref{fig:PlotFringesBPUL5U5} the interference fringes using a
$M=2$ NOON state have half the spacing compared to the single-particle case.
This gives rise to a larger slope of the probabilities as a function of
$\phi$ that, in turn, 
enables the estimation of the phase $\phi$ from the measurements
with higher sensitivity \cite{dowling_quantum_2008}. 

This argument can be made more precise by computing the quantum Fisher
information $F_Q$, which provides a lower bound on the variance of an estimator
$\hat \phi$ of the phase $\phi$ via the Cram\'er-Rao bound  
$\left( \Delta \hat \phi\right)^2\ge 1/(\nu F_Q)$, where $\nu$ is the number of
independent measurements. By the law of large numbers $\Delta \hat \phi$
decreases as $1/\sqrt{\nu}$ for increasing $\nu$. On the other hand,  
NOON states with $M$ particles
provides a quantum enhanced sensitivity for phase estimation with a variance
that decreases as $M^{-1}$. This scaling is obtained from the evaluation of 
the quantum Fisher information $F_Q$ that for pure states is
\cite{hauke_measuring_2016,mazzarella_coherence_2011}:
\begin{equation}
F_Q=4 \left(\langle \psi'(\phi)\vert \psi'(\phi)\rangle - \vert \langle
\psi'(\phi)\vert \psi(\phi)\rangle\vert^2 \right),
\label{eq:Fisher}
\end{equation}
where
 $\ket{\psi'(\phi)}=\partial \ket{\psi(\phi)}/\partial \phi$. 
 In our case the NOON state is generated by letting the initial state
 $\ket{\psi_{t=0}}\propto \left(a_1^\dag\right)^M \ket{0}$ to evolve for $t\simeq L
 U^{M-1}/J^{M}$. A relative phase factor between the endpoints is then added, as
 described in the previous section, by using a local field in the last site of the chain. 
 After these steps we get then the state  
 $\ket{\psi(\phi)}=\exp\left(-i n_L \phi \right)\ket{\psi_t}$ which, in the
 ideal case, would be a $\phi$-dependent NOON state $(\ket{M0}_{1L}+ie^{-i\phi M}
 \ket{0M}_{1L})/\sqrt2$ on sites $1,L$. 
 Therefore, in general from Eq. \eqref{eq:Fisher} we get
\begin{equation}
F_Q=4\Delta n_L^2 = 4\left(\langle n_L^2 \rangle - \langle n_L\rangle^2\right),
\label{eq:Fisher2}
\end{equation}
which in the ideal case results in $F_Q=M^2$. 
In our scheme, the ideal quantum limit can be achieved by using a fully
engineered chain \cite{banchi_perfect_2015} which enables the creation of the
ideal NOON state. 
This demonstrates the quantum enhanced
sensitivity provided by ideal NOON states. 

\begin{figure}[t]
\centering
\includegraphics[width=1\columnwidth]{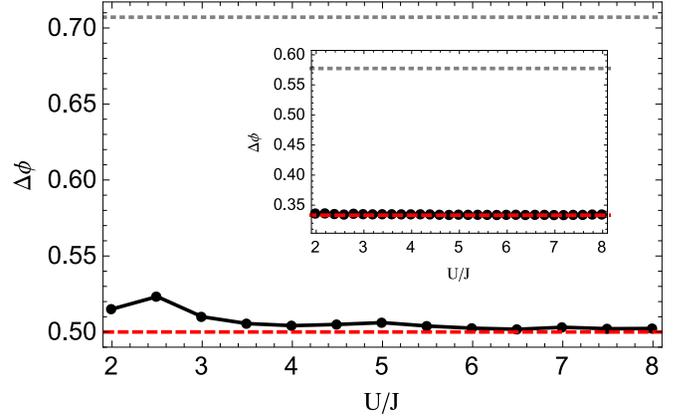}
\caption{Phase estimation precision $\Delta \phi=1/\sqrt{F_Q}$, where $F_Q$ is the Quantum Fisher Information for the estimator $n_L$ in a uniform chain with $L=5$ respectively for a two particle bound state and for a three particle bound state (inset), as a function of the onsite interaction $U/J$. The red dashed/grey dotted lines represent the ideal quantum/classical lower bound, respectively $\Delta \phi_{\mathrm{quant}}=1/M$ and $\Delta \phi_{\mathrm{cl}}=1/\sqrt{M}$.}
\label{fig:QFisher}
\end{figure}

We now show that even the imperfect NOON states obtained with uniform chains are
sufficient to achieve a quantum enhanced sensitivity. 
We consider a uniform chain with length $L=5$ and a bound state with $M={2,3}$.
In Fig. \ref{fig:QFisher} we plot the best achievable phase uncertainty $\Delta
\phi=1/\sqrt{F_Q}$, in a single measurement $\nu=1$, as a function of the onsite
interaction $U/J$ for a two and a three bound state. The grey and the red lines
represent respectively the ``classical'' limit $\Delta \phi_{\mathrm{cl}}=1/\sqrt{M}$ (obtained e.g. using coherent states where $M$ is the average number of
particles) 
and the ideal quantum limit $\Delta \phi_{\mathrm{quant}}=1/M$.
The black line on the other hand represents Eq. \eqref{eq:Fisher2}. 
Both for two and three particle bound states we observe an improvement in the phase estimation precision compared to the classical case, which is quite close to the ideal limit for an ideal NOON state and increases with the onsite interaction $U/J$. 

\section{Conclusion}
In this paper we analyze the possibility to transfer states of bound particles between the endpoints of finite lattice and their use for small cat state (NOON state) generation, using a minimal control setup.  

We derive an effective single particle theory for the dynamics of a bound particle
state in a Bose-Hubbard model with tunable couplings, using an accurate
effective Hamiltonian  technique. 
By introducing suitable static local
impurities in the edges of the lattice potential we show how to inhibit
edge-localization effects and enable the bound state dynamics. 
This allows us to realize transformations between far sites, even when strongly
interacting particles are involved. Specifically we show how state transfer
coupling schemes (in particular minimal engineering schemes), developed for
single particle states, can be introduced in our model to improve the efficiency.
We then show how to split the propagating bound state wave-function to generate
cat states (NOON states) between the endpoints of finite lattices with high
fidelity, in a minimal control setup, by tuning a single local
field in the middle of the chain.  We analyze also how environmental effects
affect our scheme, namely taking into account decoherence due spontaneous
emission in an optical lattice setup, finding the parameters' regime in which
our scheme is robust.

Our model is of interest for state transfer application with $N\ge 2$ strongly
interacting particles and for metrology application. In particular,  compared to other systems, 
it can provide some advantages for sensing external local fields in a Mach-Zehnder configuration. Indeed we
specifically show that, even in a uniform chain, the
obtained NOON states give an improvement for the phase estimation between the output arms of an interferometer. 
Moreover, our method can be straightforwardly extended to fully engineered chains 
to realize 100\% fidelity operations between distant sites, such as the perfect
state transfer of bound particle states or the perfect NOON state generation in an arbitrary long chain. 
As a future perspective, it will be interesting to adapt pumping techniques 
\cite{angelakis} or tunneling modulation techniques \cite{watanabe_efficient_2010} to speed up the transfer time, and thus 
enable the creation of higher NOON states compatible with the coherence time of the 
system. 

\section*{Acknowledgments}
The research leading to these results has received funding from the European Research Council under the European Union's Seventh Framework Programme (FP/2007-2013) / ERC Grant Agreement No. 308253. The authors thank Marco Genoni, Gerald Milburn, Hendrik Weimer and Luca Marmugi for interesting discussions and suggestions.

\appendix
\section{Effective Hamiltonian} \label{sec:EffectiveHam}
For the sake of clarity and self-consistency, here
we describe the effective Hamiltonian theory which has been derived and
successfully applied in \cite{jia_integrated_2015}. 

We assume that the Hilbert space is divided into a two subspace, the effective
subspace and the irrelevant one, which are well separated in energy when the
interaction strength $U$ is much greater than the chemical potential and the
tunneling rates.  We define a projection operator $\pr$ which projects the
states to the relevant subspace and the complementary operator
$\pq{=}\mathbf{1}-\pr$. Because $\pr$ and $\pq$ operate onto disconnected subspaces we have 
\begin{eqnarray}
\pr+\pq&=&\mathbf{1}~, \label{eq:ProjIdentity} \nonumber\\
\pr \pq &=& \pq \pr = 0~, \\
\pr^2 &=&\pr~, \nonumber\\
\pq^2 &=&\pq~. \nonumber
\end{eqnarray}
Once Eq. \eqref{eq:ProjIdentity} are applied on the Schr\"{o}dinger equation, one obtains a system of two coupled equations for the dynamics in the relevant/irrelevant subspace:
\begin{eqnarray}
i\partial_t \pr \ket{\psi}&=&\left(\pr H \pr+\pr H \pq\right)\ket{\psi}~,\\
i\partial_t \pq \ket{\psi}&=&\left(\pq H \pr+\pq H \pq\right)\ket{\psi}~.
\end{eqnarray}
Finally, using $\pr^2 {=}\pr$ and $\pq^2{=}\pq$ one obtains the system
\begin{equation}
i\partial_t 
\left(
\begin{array}{c}
\ket{\psi_p}\\
\ket{\psi_q}
\end{array}
\right)
=
\left(
\begin{array}{cc}
H_{p} & V \\
V^\dagger & H_{q}
\end{array}
\right)
\left(
\begin{array}{c}
\ket{\psi_p}\\
\ket{\psi_q}
\end{array}
\right)
\label{eq:ShroPQSystem}
\end{equation}
whence
\begin{eqnarray}
H_{p}&=&\pr H \pr~, \\
H_{q}&=&\pq H \pq~, \\
V&=&\pr H \pq~, \\
\ket{\psi_P}&=& \pr\ \ket{\psi}~,\\
\ket{\psi_Q}&=& \pq\ \ket{\psi}~.
\end{eqnarray}
Here $\ket{\psi_p}$ and $\ket{\psi_q}$ are respectively the projection of the state $\ket{\psi}$ in the relevant/irrelevant subspaces. In the interaction picture
\begin{eqnarray}
\ket{\psi_p}&=&e^{-i H_p t}\ket{\hat \phi_p}~, \\
\ket{\psi_q}&=&e^{-i H_q t}\ket{\hat \phi_q}~,
\end{eqnarray}
the free evolution is eliminated:
\begin{eqnarray}
i \partial_t\ket{\hat \phi_p}&=&e^{+i H_p t}V e^{-i H_q t} \ket{\hat \phi_q} \equiv
\hat{V}(t)\ket{\hat \phi_q}~,\\
i \partial_t \ket{\hat \phi_q}&=&e^{+i H_q t}V^\dagger e^{-i H_p t}\ket{\hat
\phi_p}\equiv\hat{V}^\dagger (t)\ket{\hat \phi_p}~,
\end{eqnarray}
We introduce $U_p$ and $U_q$ the operators that diagonalize $H_p$ and $H_q$:
\begin{eqnarray}
H_p &=& U_p  \lambda_p U^\dagger_p~, \\
H_q &=& U_q  \lambda_q U^\dagger_q~,
\end{eqnarray}
where $\lambda_p{=}\textrm{diag}\left\{\lambda_p^i\right\}$ and $\lambda_q{=}\textrm{diag}\left\{\lambda_q^i\right\}$, and defining 
\begin{eqnarray}
\ket{\tilde \phi_p}&=&U_p^\dagger \ket{\hat \phi_p}~, \\
\ket{\tilde \phi_q}&=&U_q^\dagger \ket{\hat \phi_q} ~,\\
\end{eqnarray}
we have 
\begin{eqnarray}
i\partial_t\ket{\tilde \phi_p}&=&\tilde{V}(t)\ket{\tilde \phi_q}~,\\
i\partial_t\ket{\tilde \phi_q}&=&\tilde{V}^\dagger (t)\ket{\tilde \phi_p}~,
\end{eqnarray}
and 
\begin{equation}
\tilde{V}(t)=U_p^\dagger \hat V(t) U_q=e^{i\lambda_p t}\hat V(t) e^{-i\lambda_q
t}~,
\end{equation}
Assuming that the population of the irrelevant space is initially zero $\ket{\hat \phi_q(0)}=\ket{0}$, the formal solution of system \eqref{eq:ShroPQSystem} is, in components 
\begin{equation}
\tilde{\phi}_{q,k}(t)=-i \sum\limits_{j} \int_{0}^{t}dt'\ V_{jk}^*
e^{i(\lambda_{q,k}-\lambda_{p,j})t'}  \tilde{\phi}_{p,j}(t')~,
\end{equation}
After partial integration one finds
\begin{equation}
\begin{split}
\tilde{\phi}_{q,k}(t)&=-i\sum\limits_j\left\{   
\frac{e^{i(\lambda_{q,k}-\lambda_{p.j})t'}}{i (\lambda_{q,k}-\lambda_{p.j})}V_{jk}^* \tilde{\phi}_{p,j}(t')\right|_{0}^{t}\\
&-\left. \int_{0}^{t} dt'\ \frac{e^{i(\lambda_{q,k}-\lambda_{p.j})t'}}{i (\lambda_{q,k}-\lambda_{p.j})}
\frac{d}{dt'}\tilde{\phi}_{p,j}(t')
\right\}~.
\end{split}
\end{equation}
The second integral can be neglected because carrying on the partial integration procedure the next term is of the order of $(\lambda_{q,k}-\lambda_{p.j})^{-2}$. Indeed for a large spectral separation between the relevant and the irrelevant subspaces, $\vert \lambda_Q^k-\lambda_P^j\vert{\gg}1$, and when the edge term is zero one has
\begin{equation}
\tilde{\phi}_{q,k}(t)=-\sum\limits_j \tilde{W}_{kj}(t) \tilde{\phi}_{p,j}(t)~,
\end{equation}
where
\begin{equation}
\tilde{W}_{kj}(t){=}V_{jk}^{*} \frac{\exp\left[i
\left(\lambda_Q^k-\lambda_P^j\right)t\right]}{\lambda_Q^k-\lambda_P^j}~.
\end{equation}
Finally one find that the effective Schr\"{o}dinger equation for the relevant space dynamics is: 
\begin{equation}
i\partial_t \ket{\psi_p}\simeq H_{\mathrm{eff}}\ket{\psi_p}~,
\end{equation}
where
\begin{equation}
H_{\mathrm{eff}}= H_p- V W~,
\label{eq:SCExpansion}
\end{equation}
and $W$ satisfies
\begin{equation}
H_q W - W H_p = V^\dagger. 
\label{eq:SylvesterEq}
\end{equation}
When $H_p$ consists of a degenerate levels with energy $E_0$ the above effective
theory corresponds to the usual degenerate perturbation theory 
 \cite{mila_strong-coupling_2011}: 
\begin{equation}
\elmat{\phi}{H_{\mathrm{eff}}}{\phi'}\simeq\elmat{\phi}{H}{\phi'}+\sum\limits_{m\in
    \mathcal{H}_q}\frac{\langle \phi\vert V \vert m\rangle \langle m \vert V
    \vert \phi' \rangle }{E_0-E_m}+\cdots~.
\label{eq:DegTheoryFeshbachII}
\end{equation}
where $\mathcal{H}_p$ and $\mathcal{H}_q$ are the relevant/irrelevant Hilbert subspaces and $\ket{\phi},\ket{\phi'}\in \mathcal{H}_p$.

We derive explicitly the effective Hamiltonian for a Bose-Hubbard model, Eq. \eqref{eq:BoseHubbard}. The Hilbert space $\mathcal{H}$ with fixed number of particle $M$ in a chain with length $L$ has size
\begin{equation}
\text{dim}\mathcal{H}=\frac{(M+L-1)!}{M!(L-1)!}
\end{equation}
The relevant space $\mathcal H_p\equiv \mathcal H_M$  is defined as 
\begin{equation}
\mathcal{H}_p=\left\{ \ket{\psi_m^b}\in \mathcal{H}:\ \ket{\psi_m^b}=\ket{0,\ldots, 0, M_m, 0,\ldots, 0} \right\}, 
\end{equation}
where $M_m=M$. Clearly $\text{dim}\mathcal{H}_p=L$. The irrelevant space $\mathcal{H}_q=\mathcal{H}\setminus\mathcal{H}_p$ is \begin{equation}
\begin{split}
\mathcal{H}_q=
\left\{
\ket{\psi_m^u}\in \mathcal{H}:\  \ket{\psi_m^u}=\ket{n_1,\ldots,n_L}, \right. \\
\left. \text{where } \sum_j n_j=M \text{ and } n_j\neq M  
\right\}.
\end{split}
\end{equation}
Given the above definitions we rename the basis of the Hilbert space as $\ket{m}$, $m=1,\dots,
{\rm dim} \mathcal H$ such that $\ket{m}=\ket{\psi^b_m}$ for $m=1,\dots,L$. 
The Hamiltonian then takes the following block form
\begin{equation}
H=
\left(\begin{array}{c|c}
H_p & V \\ \hline
V^\dag & H_q
\end{array}
\right), 
\end{equation}
where each block can be evaluated explicitly with the following 
projection operators 
\begin{align}
\mathbb{P}&=\sum_{m=1}^{L} \vert m\rangle\langle \psi_m^b\vert, \\
\mathbb{Q}&=\sum_{m=1}^{\text{dim}H_q} \vert L+m \rangle \langle \psi_m^u\vert. 
\end{align}
Clearly 
$\text{dim}\mathcal{H}_p=(L,L),\ \text{dim}\mathcal{H}_q=(\text{dim}\mathcal{H}-L,\text{dim}\mathcal{H}-L)$ and $\text{dim}V=(L,\text{dim}\mathcal{H}-L)$. 

The effective Hamiltonian \eqref{eq:SCExpansion} can be computed explicitly using a series expansion for large $U_j=U$. The relevant Hamiltonian for the Bose-Hubbard model 
\eqref{eq:BoseHubbard} takes the simple form 
\begin{equation}
H_p=\sum_{m=1}^{L} \frac{U}{2} M(M-1) \proj{m}{m} -\sum_{m=1}^{L} \mu_m M \proj{m}{m}.
\end{equation}
Similarly $H_q$ and $V$ can be computed explicitly. The effective model can be obtained by solving Eq. \eqref{eq:SCExpansion} and \eqref{eq:SylvesterEq}. In order to find the $W$ matrix we vectorize (see Appendix \ref{sec:Decoherence}) the Eq. \eqref{eq:SylvesterEq} as  
\begin{equation}
G \text{vec}(W)=\text{vec}(V^\dag), 
\label{eq:Gsystem}
\end{equation}
where 
\begin{equation}
G=\left(\mathbf{1}_{\text{dim}H_p}\otimes H_q\right)-\left(H_p^t\otimes \mathbf{1}_{\text{dim}H_q}\right).
\label{eq:vecSylvester}
\end{equation}
It it is convenient to write $G=G^{\text{large}}+G^{\text{small}}$ where 
\begin{align}
G^{\text{large}}&=\mathbf{1}_{\text{dim}H_p}\otimes H_q^{\text{large}}-\left(H_p^{\text{large}}\right)^t\otimes \mathbf{1}_{\text{dim}H_q},\\
G^{\text{small}}&=\mathbf{1}_{\text{dim}H_p}\otimes H_q^{\text{small}}-\left(H_p^{\text{small}}\right)^t\otimes \mathbf{1}_{\text{dim}H_p}.
\end{align}
and $H^{\text{large}}$ is the part of the Hamiltonian \eqref{eq:BoseHubbard} that contains the terms in $U$: 
$
H^{\text{large}}=\sum_{j=1}^{L} \frac{U}{2} n_j(n_j-1) \text{ and }
H^{\text{small}}=H-H^{\text{large}}
$.
The system \eqref{eq:Gsystem} can be formally solved by taking the inverse of the $G$ matrix as 
\begin{equation}
\text{vec}(W)=\frac{1}{G^{\text{large}}+G^{\text{small}}}\text{vec}(V^\dag)
\label{eq:VectorisedWEq}
\end{equation}
and using the following identity, valid for two operators $A$ and $B$:
\begin{equation}
\frac{1}{A+B}=\frac{1}{A}\left(1-B\frac{1}{A+B}\right).
\label{eq:OperatorIdentitySumInverse}
\end{equation}
Indeed one can easily find that 
\begin{equation}
\begin{split}
&\frac{1}{A}\left(1-B\frac{1}{A+B}\right)=A^{-1}(1-B (A+B)^{-1})=\\
&=A^{-1}\left[(A+B)(A+B)^{-1}-B(A+B)^{-1}\right]=\\
&=A^{-1}\left[A(A+B)^{-1}\right]=\frac{1}{A+B}.
\end{split}
\end{equation}
Using recursively the Eq. \eqref{eq:OperatorIdentitySumInverse} one finds 
the Dyson expansion
\begin{equation}
\frac{1}{G^{\text{large}}+G^{\text{small}}}=\left(G^{\text{large}}\right)^{-1}\sum_{n=0}^{+\infty} (-1)^n (G^{\text{small}}\left(G^{\text{large}}\right)^{-1})^n,
\label{eq:SerieG}
\end{equation}
that corresponds to a serie expansion in the onsite interaction parameter $U$ (which is 
contained in $G^{\text{large}}$). Moreover, 
\begin{equation}
  G^{\text{large}} = -
  \frac U2\sum_{m=1}^{L} \sum_{n=1}^{\text{dim}H_q} g(n,m)~ \ket{L{+}n,m}\bra{L{+}n,m},
  \label{eq:Glarge}
\end{equation}
where  $$ g(n,m) = \sum_j M(M-1)\delta_{jm} - n_j(n_j-1) >0, $$ since in $\mathcal H_q$ it is 
$n_j<M$ and $\sum_jn_j=M$. Therefore, $G^{\text{large}}$ is diagonal and non-singular for 
each value of $M$. 
By truncating the expansion \eqref{eq:SerieG} at the relevant order $n$ in $U$ one obtains the matrix $W$ from Eq. \eqref{eq:VectorisedWEq} and then the effective Hamiltonian from Eq. \eqref{eq:SCExpansion}, valid to the $n$-th order.

As an example of the general procedure outlined above we consider explicitly the 
first order solution where $M=2$ and  \eqref{eq:SerieG} reduces to 
\begin{align}
(G^{\text{large}}+G^{\text{small}})^{-1} &\approx (G^{\text{large}})^{-1} \\&= 
- U^{-1} \sum_{m=1}^{L} \sum_{n=1}^{\text{dim}H_q}  \ket{L{+}n,m}\bra{L{+}n,m}
\\&= - \frac{\openone_q\otimes \openone_p}{U}~.
\end{align}
Therefore, according to \eqref{eq:Gsystem}, $W=-V^\dagger/U$, so 
\begin{align}
  H_{\rm eff} &\simeq  -\sum_{m=1}^L \mu_m M \ket m \bra m + 
   \sum_{m,m'=1}^L \ket{m}\frac{\bra{\psi_m^b} V\mathbb Q V^\dagger \ket{\psi_{m'}^b}}U
  \bra{m'}, 
\end{align}
where we have explicitly omitted the terms proportional to the identity. 
For the interaction term $V$ between $\mathcal{H}_p$ and $\mathcal{H}_q$ we observe that the only non-zero matrix elements are
$\langle \psi_m^b \vert a_j a_{j+1}^\dag \vert \psi_n^u\rangle$
(as well as their Hermitian conjugate) 
when 
\begin{align}
  \ket{\psi_m^b} &=  \ket{0,\ldots,M_j,\ldots, 0} \\
  \ket{\psi_n^u} &= \ket{0,\ldots,1_j,(M-1)_{j+1},\ldots, 0}. 
\end{align}
These can give rise to a hopping from $\ket m $ to $\ket{m+1}$ only for $M=2$. Indeed,
this is done with the following steps. Starting from $\ket m = \ket{0,\dots,2_m,0\dots}$ 
the operator $a_m a_{m+1}^\dag $ in $V$ maps this state to 
$\ket{0,\dots,1_m,1_{m+1},0\dots}$ which is in $\mathcal H_q$. Then
the operator $a_m a_{m+1}^\dag$ in $V^\dagger$ maps that state to 
$\ket {m+1} = \ket{0,\dots,2_{m+1},0\dots}$. By generalizing the above 
argument, with simple calculations one finds then
\begin{align}
  H^{M=2}_{\rm eff} &\simeq  -\sum_{m=1}^L \mu_m M \ket m \bra m + 
   \sum_{m}^{L-1} \frac{J_m^2}{2U} \left(\ket{m}\bra{m+1} + {\rm h.c.}\right) . 
\end{align}

The generalization to higher values of $M$ proceeds along the same lines. For instance
for $M=3$ one has to consider the second order expansion in \eqref{eq:SerieG} which
depends also on $G^{\text{small}}$. Indeed, an effective hopping can happen only via 
a three step procedure
\begin{align}
\ket m &= \ket{0,\dots,3_m,0\dots} \to 
 \ket{0,\dots,2_m,1_{m+1}0\dots} \\&\to 
 \ket{0,\dots,1_m,2_{m+1}0\dots} \\&\to 
 \ket{0,\dots,3_{m+1}0\dots} \equiv \ket{m+1}. 
\end{align}
By doing explicit calculations we find the effective Hamiltonians mentioned in the main text.

\section{Minimal engineering of the Three Particle Bound state
propagation}\label{app:j0}
The state transfer fidelity of the three particle bound state can be improved by introducing 
an optimal coupling scheme, namely tuning the first and the last tunneling
coupling to $J_1=J_{L-1}=J_0$ and the rest of the chain to $J_j=J$. We find that,
in order to delocalize the bound state, two pairs of localized fields in the endpoints are necessary, respectively $\mu_j=-\beta_1(\delta_{j,1}+\delta_{j,L})$ and $\mu_j=-\beta_2(\delta_{j,2}+\delta_{j,L-1})$ where $\beta_1=(2J^2-J_0^2)/8U$ and $\beta_2=(J^2-J_0^2)/8U$. In this case the beam splitting condition for $L\gg 1$ is realized when a local field $\mu_j=-\beta\delta_{j,L/2+1}$ with strength $\beta=\tilde{\beta} J^3/8U^2$ is added in the middle of the chain. The effective Hamiltonian is it in this case 
\begin{equation}
 H_{\mathrm{opt}}^{III}/2J_{\mathrm{eff}}^{III}=
\left(
\begin{array}{ccccccc}
U_{\mathrm{opt}}^{III}& \frac{J_0^3}{2 J^3} &   &   &   &   &   \\
 \frac{J_0^3}{2 J^3} & \ddots & 1/2 &   &   &
     &   \\
   & 1/2 & U_{\mathrm{opt}}^{III}& \ddots &   &   &   \\
   &   & \ddots & U_{\mathrm{opt}}^{III}+\tilde{\beta}& \ddots &   &   \\
   &   &   & \ddots &U_{\mathrm{opt}}^{III}  & 1/2 &   \\
   &   &   &   & 1/2 & \ddots & \frac{J_0^3}{2
   J^3} \\
   &   &   &   &   & \frac{J_0^3}{2 J^3} & U_{\mathrm{opt}}^{III}\ 
\end{array}
\right)
\end{equation}
where $J_{\mathrm{eff}}^{III}=3J^3/16 U^2$, 
$U_{\mathrm{opt}}^{III}= 8 (U/J)^3+2 U/J+16 (J/U)^2 $. In order to have a perfectly
balanced beam splitter in this single-particle Hamiltonian, as shown in 
\cite{compagno_toolbox_2015}, one needs $\tilde\beta=1$. Therefore,
$\beta^{50/50}{=}J^3/8U^2$. 
Our method is straightforwardly generalizable to bound states with a higher
number of particles. 

\section{Numerical solution of the master equation} \label{sec:Decoherence}
In order to solve the Master equation \eqref{eq:MasterEqDecoherence} we exploit
a vectorization procedure \cite{am-shallem_three_2015} which consists of representing a matrix as a vector, by using its representation in the canonical basis with a column ordering. For instance for a generic $2$x$2$ matrix $A$, its vectorization $\mathrm{vec}(A)$ is 
\begin{equation}
A=
\left(
\begin{array}{cc}
A_{11} & A_{12} \\
A_{21} & A_{22}
\end{array}
\right)
=
A_{11} 
\left(
\begin{array}{cc}
1&0\\
0&0
\end{array}
\right)
+
A_{21}
\left(
\begin{array}{cc}
0&0\\
1&0
\end{array}
\right)
+\ldots 
\end{equation}
\begin{equation}
\mathrm{vec}(A)=
\left(
\begin{array}{cccc}
A_{11},&A_{21},&A_{12},&A_{22}
\end{array}
\right)^t~.
\label{eq:Vectorisation}
\end{equation}
For a general  size matrix $\rho$ the latter procedure corresponds to the mapping $v_{(k-1)L+j}=\rho_{j k}$ where $v=\mathrm{vec}(\rho)$. Once chosen this base the action of an operator $H$ on the left or the right of the density matrix $\rho$ can be written as
\begin{eqnarray}
H\rho&=&\left(\mathbf{1}_{L}\otimes H\right)\mathrm{vec}(\rho)~,\\
\rho H&=&\left( H^t \otimes \mathbf{1}_{L}\right)\mathrm{vec}(\rho)~,
\end{eqnarray}
and for the dissipative part, by using the identities 
\begin{eqnarray}
\mathrm{vec}(ABC) &=&\left(C^t\otimes A\right) \mathrm{vec}(B)=\left(\mathbf{1}_L
\otimes AB\right) \mathrm{vec}(C)=\\ \nonumber
&=&\left(C^t B^t \otimes \mathbf{1}_L \right) \mathrm{vec}(A)~,\\
\mathrm{vec}(AB)&=&\left( \mathbf{1}_L \otimes A\right) \mathrm{vec}(B)=\left( B^t
\otimes \mathbf{1}_L\right) \mathrm{vec}(A)~,
\end{eqnarray}
we find that
\begin{eqnarray}
n_i n_i \rho &=&\left(\mathbf{1}_{L}\otimes n_i n_i\right)\mathrm{vec}(\rho)~,\\
\rho n_i n_i&=&\left( (n_i n_i)^t \otimes \mathbf{1}_{L}\right)\mathrm{vec}(\rho)~,\\
n_i \rho n_i&=&\left( (n_i )^t \otimes n_i \right)\mathrm{vec}(\rho)~,
\end{eqnarray}
hence if $H$ and $\rho$ describe a fixed particle number subspace one obtains the vectorised version of the master equation 	\eqref{eq:MasterEqDecoherence} namely
\begin{equation}
\mathrm{vec}(\dot{\rho})=\mathcal{L}_v\mathrm{vec}(\rho)~,
\end{equation}
where the operator $\mathcal{L}_v$ is defined as 
\begin{eqnarray}
&\mathcal{L}_v=&-i\left\{\left(\mathbf{1}_L \otimes H\right)-\left(H^t\otimes \mathbf{1}_L \right)\right\}-\gamma\sum\limits_j \left\{\left(n_j^t\otimes n_j
    \right)+\right. \\ \nonumber
&&-\left. \frac{1}{2}\left(\mathbf{1}_L\otimes n_j^2\right)-\frac{1}{2}\left((n_j^2)^t\otimes \mathbf{1}_L\right)\right\}.
\end{eqnarray}

\end{document}